\documentclass[twocolumn]{aastex63}

\usepackage{graphicx}
\usepackage{subfigure}
\usepackage{amsmath}
\usepackage{amssymb}
\usepackage{booktabs}
\usepackage{indentfirst}
\usepackage{verbatim}
\usepackage{array}

\usepackage{newtxtext,newtxmath}
\usepackage{ae,aecompl}

\newcommand {\dindex}{{\rm D_{n}(4000)}}
\newcommand {\hdeltaf}{{\rm H\delta_{\mathrm F}}}

\newcommand {\rtf} {R_{\mathrm 25}}
\newcommand {\rd} {R_{\mathrm d}}
\newcommand {\re} {R_{\mathrm e}}
\newcommand {\rbreak} {R_{\mathrm {break}}}
\newcommand {\Ms} {M_{\star}}
\newcommand {\TI} {\sc Ti}
\newcommand {\TII} {\sc Tii}
\newcommand {\TIII} {\sc Tiii}

\submitjournal{ApJ}

\shorttitle{Surface brightness profile breaks of disk galaxies from MaNGA}
\shortauthors{Tang, Chen and Zhang et al.}

\begin{document}

\title{New Constraints On the Origin of Surface Brightness Profile Breaks of Disk Galaxies From MaNGA}

\correspondingauthor{Hong-Xin Zhang, Xu Kong}
\email{hzhang18@ustc.edu.cn, xkong@ustc.edu.cn}

\author{Yimeng Tang}
\altaffiliation{co-first authors}
\author{Qianhui Chen}
\altaffiliation{co-first authors}
\author{Hong-Xin Zhang}
\author{Zesen Lin}
\author{Guangwen Chen}
\author{Yulong Gao}
\author{Zhixiong Liang}
\author{Haiyang Liu}
\author{Xu Kong}
\affiliation{Key Laboratory for Research in Galaxies and Cosmology, Department of Astronomy, \\University of Science and Technology of China, Hefei 230026, China}
\affiliation{School of Astronomy and Space Science, University of Science and Technology of China, Hefei 230026, China}

\begin{abstract}

In an effort to probe the origin of surface brightness profile (SBP) breaks widely observed in nearby disk galaxies, we carry out a comparative study of stellar population profiles of 635 disk galaxies selected from the MaNGA spectroscopic survey.\ We classify our galaxies into single exponential ({\TI}), down-bending ({\TII}) and up-bending ({\TIII}) SBP types, and derive their spin parameters and radial profiles of age/metallicity-sensitive spectral features.\ Most {\TII} ({\TIII}) galaxies have down-bending (up-bending) star formation rate (SFR) radial profiles, implying that abrupt radial changes of SFR intensities contribute to the formation of both {\TII} and {\TIII} breaks.\ Nevertheless, a comparison between our galaxies and simulations suggests that stellar migration plays a significant role in weakening down-bending $\Sigma_{\star}$ profile breaks.\ While there is a correlation between the break strengths of SBPs and age/metallicity-sensitive spectral features for {\TII} galaxies, no such correlation is found for {\TIII} galaxies, indicating that stellar migration may not play a major role in shaping {\TIII} breaks, as is evidenced by a good correspondence between break strengths of $\Sigma_{\star}$ and surface brightness profiles of {\TIII} galaxies.\ We do not find evidence for galaxy spin being a relevant parameter for forming different SBP types, nor do we find significant differences between the asymmetries of galaxies with different SBP types, suggesting that environmental disturbances or satellite accretion in the recent past do not significantly influence the break formation.\ By dividing our sample into early and late morphological types, we find that galaxies with different SBP types follow nearly the same tight stellar mass-$\rtf$ relation, which makes the hypothesis that stellar migration alone can transform SBP types from {\TII} to {\TI} and then to {\TIII} highly unlikely.\

\end{abstract}

\keywords{galaxies: formation -- galaxies: evolution -- galaxies: stellar content -- galaxies: structure -- galaxies: spiral -- galaxies: general}

\section{Introduction}

The radial surface brightness profiles of disk galaxies have been studied since the middle of the last century, and it was once thought that they follow a single exponential decline 
(\citealt{Patterson1940,deVaucouleurs1959,Freeman1970}). However, since the seminal work by \citet{vanderKruit1979}, it has been widely recognized that, instead of following 
single exponential declining profiles, a large number of galaxies have sharp truncations or breaks in their surface brightness profiles. By analyzing CCD imaging data of large samples 
of nearby disk galaxies, \citet{Erwin2005} and \citet{Pohlen2006} divided disk galaxy surface brightness profiles into three main types: single exponential profiles (Type I, hereafter 
{\sc Ti}), down-bending double exponential profiles (Type II, hereafter {\sc Tii}) and up-bending double exponential profiles (Type III, hereafter {\sc Tiii}). \cite{Pohlen2006} found that 
nearly 90\% of their spiral galaxies have broken surface brightness profiles, of which 60\% are down-bending profiles and 30\% are up-bending profiles.\ A minor fraction of galaxies 
even have more than one breaks.\ Late-type dwarf galaxies continue the trend established for spiral galaxies, with a much larger fraction of {\sc Tii} than {\sc Ti} and {\sc Tiii} profiles 
\citep{Herrmann2013}.\ Observations of high redshift galaxies suggest that surface brightness profile breaks are already present in the early universe (\citealt{Perez2004,Azzollini2008}).

{\sc Tii} profiles are the most common ones among the three disk types.\ Various mechanisms have been proposed to explain the origin of such down-bending surface brightness 
profiles.\ The different mechanisms may be broadly sorted into three categories.\ The first category attributes the breaks to the maximum angular momentum of the protogalactic clouds 
which collapsed to form the present-day galaxies \citep{vanderKruit1987,vanderKruit1988}, the second category invokes an abrupt change in star formation radial profiles (e.g.,\ \citealt{Kennicutt1989,
Schaye2004,Elmegreen2006}), and the third category invokes stellar radial migration induced by resonant scattering of bars and/or spiral arms (e.g.,\ \citealt{Sellwood2002,Debattista2006,
Roskar2008a,Minchev2012,DiMatteo2013}).\ The first category of mechanisms predicts a break radius of around 4 to 5 times the radial scalelengths, which is not consistent with the 
general observational results that {\sc Tii} break radii are $\lesssim$ 3 times the scalelengths \citep[e.g.,][]{Pohlen2006}.\ Stellar migration has been widely advocated to explain the 
U-shaped color/age profiles together with nearly absent stellar mass surface density profile breaks observed in many {\sc Tii} galaxies \citep[e.g.,][]{deJong2007,Bakos2008,
Yoachim2012,Radburn-Smith2012}.\ Nevertheless, simulations within a fully cosmological context found that an abrupt radial change in star formation efficiencies can naturally form 
down-bending disk profiles, while secular radial migration may play a role in weakening the magnitude of disk profile breaks \citep{Martinez-Serrano2009, Sanchez-Blazquez2009}.\

{\sc Tiii} profiles are the most commonly observed type in early-type galaxies \citep[][]{Erwin2005, Gutierrez2011}.\ \citet{Bakos2008} find that galaxies with up-bending surface 
brightness profiles usually also have up-bending stellar mass surface density profiles.\ Like the {\sc Tii} profiles, enhanced star formation efficiencies either within or beyond the break radii have been invoked to explain {\sc Tiii} breaks in actively star-forming galaxies (\citealt{Hunter2006,Laine2016,Wang2018}).\ \citet{Borlaff2018} find that the break location and scaling relations of {\sc Tiii} S0 galaxies up to redshift of 0.6 are compatible to that of their nearby counterparts, and thus suggest that formation of up-bending profiles might be related to gravitational and dynamical processes, contrary to the formation mechanisms proposed for down-bending profiles.\ It was also suggested that some {\sc Tiii} profiles might be an artifact caused by a superposition of a smaller single-exponential thin disk and a larger single-exponential thick disk \citep{Comeron2012}.\ Recent simulations by \cite{Herpich2017} suggest that a strong bar may drive stellar radial migration in low-spin galaxies which leads to formation of up-bending profiles.\ Nevertheless, environmental influences are probably the most widely advocated formation mechanisms for up-bending profiles (e.g.,\ \citealt{Erwin2005,Laine2014,Watkins2019}).\ In particular, \cite{Younger2007} show that minor mergers can drive gas inflow toward galaxy centers and at the same time outward transfer of angular momentum, which may result in the formation of up-bending disk profiles.\ Similarly, \cite{Ruiz-Lara2017} suggest that satellite accretion may help accumulate mass in outer disks and produce up-bending disk profiles.\ Moreover, \citet{Borlaff2014} show that major mergers can also produce S0-like remnants with up-bending profiles.\ 

Several studies have attempted to explain different disk profile types under a common framework.\ Recent simulations by \citet{Herpich2015} suggest that there is a correlation 
between dark matter halo's initial angular momentum and the resulting stellar radial profiles.\ In particular, galaxies living in the lowest spin haloes show up-bending disk profiles, 
while galaxies living in the highest spin haloes develop down-bending disk profiles.\ In addition, by invoking the two mechanisms of secular stellar migration and satellite accretion, 
\citet{Ruiz-Lara2017} show that the {\sc Ti} profile can be an intermediate and transitional type from {\sc Tii} to {\sc Tiii} profiles, with the {\sc Tiii} ({\sc Tii}) galaxies subject to the 
strongest (weakest) influence of stellar migration and satellite accretion.\ 

With the advent of large integral field unit (IFU) spectroscopic surveys, such as the Calar Alto Legacy Integral Field Area survey (CALIFA; \citealt{Sanchez2012}), 
Sydney-Australian-Astronomical-Observatory Multi-object Integral-Field Spectrograph Galaxy Survey (SAMI; \citealt{Croom2012}) and Mapping Nearby Galaxies 
at Apache Point Observatory survey (MaNGA; \citealt{Bundy2015}), it has become possible to obtain relatively robust constraints on stellar population distribution in 
large samples of nearby galaxies, which is crucial for a straightforward test of various disk profile formation mechanisms.\ Based on the CALIFA data, \cite{Marino2016} 
and \cite{Pilyugin2017} studied the radial gradients of nebular gas abundances of galaxies with different disk profile types.\ \cite{Ruiz-Lara2016} carried out full spectrum 
stellar population fitting to 44 CALIFA galaxies and found that U-shaped stellar age radial profiles are present in both {\sc Ti} and {\sc Tii} galaxies.\

In an effort to gain further insight into the formation mechanisms of different types of disk profiles, the current work makes use of the available Sloan Digital Sky Survey IV 
(SDSS-IV) MaNGA data to perform a comparative study of stellar population radial variations, galaxy spin and morphologies of galaxies with different profile types.\ The rest 
of this paper is structured as follows.\ Section \ref{sec:data} presents our sample selection.\ Section \ref{sec:method} describes the method used in our data analysis.\ The 
main results from our analysis are given in Section \ref{sec:results}.\ A summary and discussion follow in Section \ref{sec:summary}.

\section{Data and Sample}\label{sec:data}

\subsection{Data}

The SDSS-IV MaNGA project has been collecting fiber-bundle IFU spectroscopic observations of a representative sample of approximately 10,000 galaxies 
in the redshift range of $0.01 < z < 0.15$.\ The MaNGA sample is selected from an extended version of the NASA-Sloan Atlas catalogue \citep{Blanton2011} and have 
a nearly flat distribution in stellar mass from $\sim$ 10$^{9}$ to 3$\times$10$^{11}$ $M_{\odot}$ \citep{Wake2017}.\ MaNGA uses a series of hexagonal optical fiber 
bundles with different sizes in order to cover galaxies out to 1.5 $r$-band effective radii ($R_{e}$) for the Primary sample and to 2.5 $R_{e}$ for the Secondary sample.\
By using the dual-channel BOSS spectrographs \citep{Smee2013}, MaNGA achieves a continuous wavelength coverage from 3600 {\AA}~to 10300 {\AA}~with a resolving 
power of $R$ $\simeq$ 2000.\ The reduced datacubes \citep{Law2016} have a spaxel size of 0\farcs5 and a typical effective spatial resolution of FWHM $\sim$ 2\farcs5 
\citep{Law2015}.\ With roughly 3-hr dithered exposures per field, MaNGA reaches a typical S/N of $8.3$ ({\AA$^{-1}$} per fiber) at $\sim$ 1.5$R_{e}$ and $2.3$ ({\AA$^{-1}$} 
per fiber) at $\sim$ 2.5$R_{e}$ (\citealt{Wake2017}).\

The latest SDSS DR15 (\citealt{Aguado2019,Fischer2019}) includes MaNGA observations for 4621 galaxies, among which 2956 are from the MaNGA Primary 
sample and 1665 are from the Secondary sample.\ We work with galaxies of the Secondary sample in this paper in order to reach beyond the typical radii of surface brightness 
breaks in nearby disk galaxies ($\sim$1.5 $R_e$, \citealt{Marino2016}).\ In addition to the reduced spectral datacubes, DR15 also includes higher-level data products (e.g.,\ stellar 
and emission-line velocity field) produced by the MaNGA Data Analysis Pipeline (DAP, \citet{Westfall2019}).\ These higher-level data products will be used in the following manipulation 
of MaNGA spectra in this work.
{
We use the SDSS $g$- and $r$-band images to derive surface brightness profiles of our spectroscopic sample.\ The surface brightness profiles will be used to classify our galaxies into different disk profile types.

\subsection{Sample selection}\label{sec:sample}
{
In order to select disk galaxies from the MaNGA Secondary sample, we make use of the MaNGA Deep Learning Morphology Value Added Catalogue 
\citep[MDLM-VAC,][]{Dominguez-Sanchez2018, Fischer2019} which provides Deep Learning based morphological classifications for the SDSS DR15 MaNGA sample.\ 
Following \cite{Fischer2019}, we use the T-Type \footnote{\hspace{-1pt}\textbf{The MDLM-VAC T-Types are obtained by training with the T-Type catalog of \cite{Nair2010} which follows the same morphological classification scheme as RC3 \citep{1991rc3..book.....D} for S0 and later types except that S0+ and S0 are assigned the same T-Type value of $-2$.\ The correspondence between T-Types and Hubble morphological types is indicated in the left panel of Figure \ref{fig:disktype}.\ Note that, unlike the integer T-Type values in \cite{Nair2010},  MDLM-VAC T-Types are floating point numbers.}} and $P_\mathrm{S0}$ parameters from MDLM-VAC to select spiral (T-Type $>$ 0) and S0 (T-Type $\leqslant$ 0, $P_{S0}>0.5$) 
galaxies from the Secondary sample.\ The $P_\mathrm{S0}$ parameter quantifies the probability of a galaxy being S0 when T-Type $\leqslant$ 0.\ In addition, we also require the 
apparent minor-to-major axis ratio $b/a>0.5$ (determined based on our isophotal analysis described below) in order to avoid very edge-on galaxies.\ The above selection criteria 
result in 885 relatively face-on spiral or S0 galaxies.\ 

We perform a visual inspection of SDSS images of the above-selected 885 galaxies, and find that 83 of them appear to be either involved in interacting/merging 
event or substantially contaminated by bright foreground stars.\ These 83 galaxies are excluded from our following analysis.\ In addition, as we will show in Section 
\ref{sec:sbpfit}, 18 of the remaining 802 galaxies turn out to have radial surface brightness profiles that are best fitted by non-exponential S\'ersic functions 
(i.e.,\ S\'ersic index $\neq$ 1), and another 61 are best fitted by piece-wise exponential functions with more than one breaks.\ These 79 galaxies are excluded from 
the current work.\ Lastly, 88 of the remaining 723 galaxies have surface brightness profile break radii (see Section \ref{sec:sbpfit}) that fall outside of the spatial 
coverage of MaNGA IFU observations, so these galaxies are excluded from further analysis.\ Therefore, our final sample includes a total of 635 galaxies.\
}

\section{Data analysis}\label{sec:method}

\subsection{Derivation of radial surface brightness profiles}

\begin{figure*}
\centering
\includegraphics[width=1\textwidth]{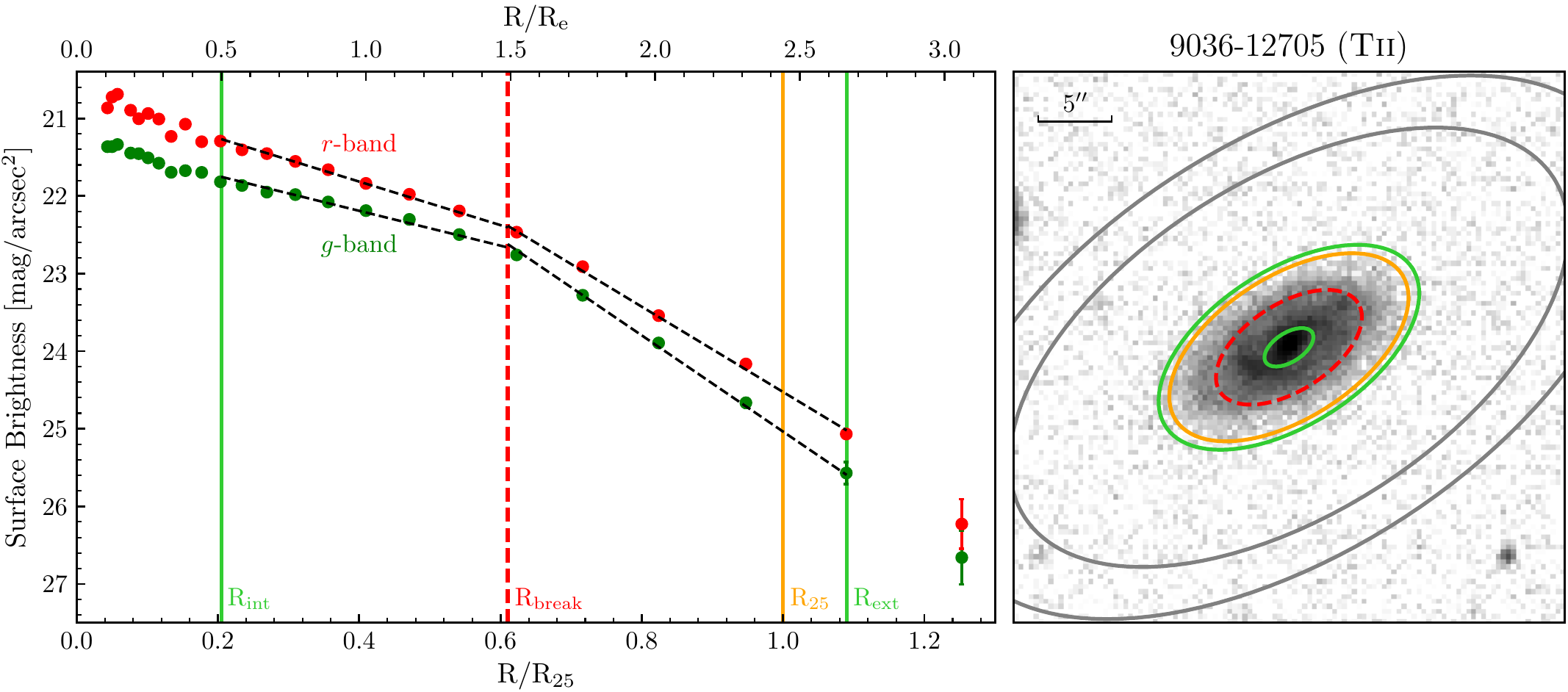}
\caption{
An example (MaNGA ID: 9036-12705) of the surface photometry performed on SDSS images.\
The left panel shows the derived $g$- and $r$-band surface brightness profiles, and the right panel shows the $g$-band image.\ The black dashed lines in the left 
panel represent the best-fit double exponential profiles.\ The surface brightness profile break radius ($\rbreak$), $g$-band 25 mag arcsec$^{-2}$ isophotal radius 
($\rtf$) and the inner/outer radial bounds for profile fitting ($R_{\rm int}$ and $R_{\rm ext}$) are indicated in both the left (as straight lines) and right (as ellipses) panels, 
following the same color scheme and line styles.\ The two outermost gray ellipses overplotted on the image mark the radial range used for sky background estimation.
}
\label{fig:surfphot}
\end{figure*}

We perform surface photometry on the SDSS $g$- and $r$-band images of galaxies in the MaNGA Secondary sample using the Image Reduction and Analysis 
Facility (IRAF) task {\sc ellipse}.\ To eliminate contamination by foreground and background sources, we create IRAF pixel mask files based on the SEGMENTATION 
map generated by {\sc SExtractor} \citep{Bertin1996}.\ We use the galaxy centers, position angles (PA) and ellipticities from the MaNGA PyMorph Photometric Value 
Added Catalog \citep{Fischer2019} as the initial geometric parameters for running {\sc ellipse}.\ The isophotal fitting with {\sc ellipse} is carried out in two steps.\ 
In the first step, the galaxy center, PA and ellipticity are allowed to vary with radius in the fitting.\ In the second step, average values of the best-fit geometric parameters 
around 1.5$R_{e}$ determined in the first step are adopted and kept constant in the second {\sc ellipse} run.\ 

To determine local sky background for each galaxy, we use a 15-pixel wide elliptical annulus at 5$R_{e}$ from the galaxy center and divide the annulus into 15 sectors 
of equal area.\ The sky level and its standard deviation are calculated as the median and standard deviation of median pixel values of the 15 sky sectors.\ The total error 
budget of our photometry is dominantly contributed by the Poisson noise and uncertainties in sky subtraction.\ As an example, Figure \ref{fig:surfphot} shows the SDSS 
image, surface brightness profiles and exponential profile fitting (see Section \ref{sec:sbpfit}) for one galaxy selected from our sample.

\subsection{Surface brightness profile fitting}\label{sec:sbpfit}
We fit the $r$-band radial surface brightness profiles with four sets of model functions.\ These model functions are, in order of increasing complexity, single exponential,
single S\'ersic, double exponentials and triple exponentials.\ The model fitting is limited to data points that are more than 3-$\sigma$ brighter than the background.\ 
We use the Levenberg-Marquardt least-squares minimization algorithm implemented in MPFIT to find the best-fit parameters for each of the four model functions, and 
then adopt the corrected Akaike information criterion (AICc) to select the simplest model function that gives an adequate fit to the radial profile of each galaxy.\ As a sanity 
check, we also perform a visual inspection of the residuals of model fitting as a function of radius, and find that the AICc-selected best models for $\sim$ 10\% of our 
galaxies do not yield significantly smaller residuals with radius than the next best models with fewer parameters.\ So we revise the model selection accordingly for these 
galaxies.\

Because our interest in this work is to study disk galaxies with either single or double exponential profiles, we exclude 18 galaxies that are best fitted with non-exponential 
S\'ersic models (i.e.,\ S\'ersic index $\neq$ 1) and another 61 galaxies that are best fitted with triple exponential models from the following analysis.\ The 61 triple-exponential 
disk galaxies will be the subject of our next work.\ To eliminate the influence of galaxy bulges on disk profile fitting for the remaining 723 galaxies, we first make a visual 
identification, if any, of the transition radius from bulge- to disk-dominated regions, and then repeat the exponential model fitting by continuously adjusting the inner radial bound 
for profile fitting around the visually-identified transition radius until the scale length of the best-fit (inner) exponential model becomes stable.\ 

With the surface brightness profile fitting, our final sample is classified into single exponential profiles ({\sc Ti}) and double (broken) exponential profiles.\ The broken exponential 
profiles are further classified into down-bending profiles whose outer exponentials have steeper radial declining than the inner exponentials ({\sc Tii}), and up-bending 
profiles whose outer exponentials have shallower radial declining than the inner exponentials ({\sc Tiii}).

\subsection{Spectral stacking}\label{sec:stacking}
This work aims to exploit radial gradients of age- or metallicity-sensitive spectral features, including nebular emission lines and Lick absorption line indices 
\citep{Worthey1994,Worthey1997}, to probe the formation mechanisms of disk breaks.\ In order to obtain robust measurements of the spectral features, 
especially the Lick absorption line indices which generally requires spectral S/N $\gtrsim$ 20-30 \AA$^{-1}$, we perform spectral stacking (by summing 
individual spectra) as a function of galactocentric radius.\ Given the difference between emission line velocity field and absorption line velocity field, two 
sets of spectral stacking are carried out, one for measuring emission lines and the other one for Lick absorption line indices.\ 

Individual spectra are corrected for Galactic extinction using the \cite{Schlegel1998} extinction map, degraded to the wavelength-dependent Lick/IDS spectral 
resolution, and then shifted to rest frame before stacking, by using the DAP emission line and absorption line velocity field, respectively, for measuring emission 
and absorption features.\ The geometric parameters used for stacking spaxels as a function of radius are the same as those derived from our $r$-band isophotal 
analysis.\ The stacking runs from the center to larger radii of each galaxy, with the width of contiguous and non-overlapping radial bins for stacking iteratively increasing 
outward until the stacked spectra reach S/N $\geqslant$ 30 \AA$^{-1}$~at wavelength $\sim$ 5500 \AA.\ Only spaxels with spectral S/N $>$ 2 are used in our stacking.\

\subsection{Measurements of spectral features and derivation of stellar mass surface densities}

\begin{deluxetable*}{ccccc}
\tabletypesize{\small}
\tablecolumns{5}
\setlength{\columnsep}{0.002pt}
\tablewidth{0pt}
\tablehead{
\colhead{Lick index} 
&\colhead{Blue continuum}
&\colhead{Feature}
&\colhead{Red continuum}
&\colhead{Units}
}

\startdata
H$\delta_{\mathrm{F}}$ & 4057.250-4088.500 & 4091.000-4112.250 & 4114.750-4137.250 & \AA \\
Fe4383       & 4359.125-4370.375 & 4369.125-4420.375 & 4442.875-4455.375 & \AA \\
Mg$_1$       & 4895.125-4957.625 & 5069.125-5134.125 & 5301.125-5366.125 & mag \\
Mg$b$        & 5142.625-5161.375 & 5160.125-5192.625 & 5191.375-5206.375 & \AA \\
Fe5270       & 5233.150-5248.150 & 5245.650-5285.650 & 5285.650-5318.150 & \AA \\
Fe5335       & 5304.625-5315.875 & 5312.125-5352.125 & 5353.375-5363.375 & \AA \\
\enddata

\caption{Lick indices studied in this work, following the definition by \citet{Worthey1994} and \citet{Worthey1997}.}
\label{tab:lickindex}
\end{deluxetable*}

Based on the stacked spectra, we are interested in measuring hydrogen recombination emission lines, the 4000-\AA~break D$_{n}$(4000) \citep{Balogh1999} and 
Lick absorption line indices \citep{Worthey1994,Worthey1997}.\ An accurate measurement of the spectral features, particularly the Balmer lines, requires a proper 
decomposition of the observed spectrum into nebular emission lines and pure stellar absorption spectrum.\ To this end, we use the Penalized Pixel Fitting (pPXF, 
\citealt{Cappellari2012}) software to perform a simultaneous fit of stellar population and Gaussian emission line templates to the stacked spectra.\ For the stellar population 
models, we use single stellar population models covering 15 different ages from 0.063 Gyr to 15.8 Gyr and 6 different metallicities from [\emph{Z}/H] $=$ $-1.71$ to 0.22.\ 

The Balmer emission line flux is derived directly from the best-fit Gaussian emission-line templates, and the emission line equivalent widths (EW) are obtained by dividing 
the line flux by local pseudo-continua of the observed spectra.\ The absorption line features are measured on the emission line subtracted spectra.\ In addition to measuring 
the spectral features, we also use the stellar mass-to-light ratios returned by pPXF, together with the measured $r$-band surface photometry, to derive stellar mass surface 
density $\Sigma_{\star}$ profiles of our galaxies.

\subsection{Derivation of radial profiles of age/metallicity sensitive spectral indices and star formation}
With the above measured spectral indices in hand, we choose to focus on radial profiles of the composite index [MgFe$]^{\prime}$ introduced by \cite{Thomas2003} 
which traces the total metallicities, the magnesium to iron index ratio Mg$/$Fe which primarily traces the $\alpha/$Fe ratio and (thus) star formation timescales, the 
H$\delta_{\mathrm{F}}$ index which is among the most age-sensitive Lick indices, and D$_{n}$(4000) which serves as an extinction-free indicator of stellar ages and, 
especially at old ages, metallicities.\ We also use the extinction corrected (with Balmer decrement method) H$\alpha$ luminosities to trace the star formation rate (SFR) 
averaged over the recent $\sim 10-20$ Myr.\

Definition of the Lick indices \citep{Worthey1994,Worthey1997} studied in this work is given in Table \ref{tab:lickindex}.\ The composite index [MgFe$]^{\prime}$ is a 
combination of the Mg$b$, Fe5270 and Fe5335 indices and is defined as 
\begin{eqnarray}
[\mathrm{MgFe}]^{\prime}=\sqrt{{\rm Mg}b(0.72\times{\rm Fe5270}+0.28\times{\rm Fe5335})}
\end{eqnarray}
and the [Mg$/$Fe] index is represented by Mg$b/\langle$Fe$\rangle$ in \cite{Thomas2003}, where $\langle$Fe$\rangle$ $=$ (Fe5270 $+$ Fe5335)$/2$.\ A  
majority of the MaNGA spectra exhibit noticeable sky-subtraction residuals near the observed wavelength of O{\sc I} 5577 sky emission line, which results in a 
contamination to the redshifted Mg$b$, Fe5270 and Fe5335 features, respectively, for 7.5\%, 20\% and 36.5\% of our galaxies.\ Therefore, we decide to use the 
Fe4383 index as a substitute for Fe5270 and Fe5335 for our whole sample.\ \cite{Thomas2003} shows that the Fe4383 index traces Fe abundance nearly as well 
as the classic indices Fe5270 and Fe5335.\ For Mg indices, we use the Mg$_1$ index as a substitute for the 7.5\% galaxies whose Mg$b$ features are contaminated 
by the sky-subtraction residuals.\ \cite{Thomas2003} shows that the response of Mg$_1$ to $\alpha$ element abundance changes is very similar to Mg$b$.\

In order to keep a formal consistency with the original definition of the composite indices, we determine the following empirical relationships based on our galaxies with none 
of the above mentioned Lick absorption features affected by sky-subtraction residuals:
\begin{eqnarray}
\mathrm{Fe5270}=0.35 \mathrm{Fe4383}+0.96 \\
\mathrm{Fe5335}=0.30 \mathrm{Fe4383}+1.23 \\
\mathrm{Mg}b=26.15 \mathrm{Mg_1}+1.16
\end{eqnarray}
These relations have r.m.s.\ scatters $\leqslant$ 0.2.\ 
We follow the definition in \cite{Thomas2003} to derive [MgFe$]^{\prime}$ and Mg$/$Fe, by using the indices Fe5270, Fe5335 and Mg$b$ inferred based on the above 
three equations.

\subsection{Classification of the radial profiles of $\Sigma_{\star}$ and $\hdeltaf$}\label{sec:classspectr}
Similar to the practice of classifying surface brightness profiles into {\sc Ti}, {\sc Tii} and {\sc Tiii} (Section \ref{sec:sbpfit}), we perform linear regression on the $\log$$\Sigma_{\star}$ and 
$\hdeltaf$ radial profiles and then make a classification of their basic radial trends.\ Specifically, we classify $\Sigma_{\star}$ radial profiles into single exponential, down-bending 
double exponential and up-bending double exponential declining profiles, in analogy to the {\sc Ti}, {\sc Tii} and {\sc Tiii} surface brightness profile types.\ We classify the $\hdeltaf$ profiles into 
single linear, up-bending, down-bending-D, down-bending-F and down-bending-$\Lambda$ radial shapes, where down-bending-D means the radial gradient becomes shallower 
but do not change direction beyond the break radius, down-bending-F means the radial gradient becomes flat (i.e., being consistent with zero within 2$\sigma$ uncertainties) 
beyond the break radius and down-bending-$\Lambda$ means an approximately $\Lambda$-shaped radial profile.\ 

We point out that spectral-index radial profiles can not always be accurately represented by either single or broken linear relations, and our intention here is just to capture the 
basic radial trend.\ The $\Sigma_{\star}$ profile classification will be used for exploring the influence of galaxy spin on the formation of stellar radial profile breaks.\ The $\hdeltaf$ 
profile classification will be used as a proxy for stellar age profile classification.\ Among the three subtypes of down-bending $\hdeltaf$ profiles, the $\Lambda$-shape profiles 
correspond to the familiar U-shape age profiles which have been invoked as an observational evidence for stellar radial migration effect.

\subsection{Derivation of the spin parameter}\label{sec:method_spinpars}
Some recent simulations suggest that the halo spin parameter may play an important role in shaping different disk profile types.\ Generally speaking, the halo spin parameter 
$\lambda$ can not be strictly calculated based on observations.\ \cite{Hernandez2007} developed a method to obtain an approximate estimate of halo spin parameters of disk 
galaxies based on the observed stellar disk size, as quantified by the exponential disk scalelength, and maximum stellar rotation velocity (Equation 9 in \citealt{Hernandez2007}).\ 
We choose to use the \cite{Hernandez2007} method to estimate spin parameters of our galaxies.\

Most of our galaxies have double-exponential radial profiles, so it is not straightforward to define an exponential disk scalelength for $\lambda$ estimation.\ To proceed, we first 
derive the disk half-light radius $R_{\rm e, disk}$ for each galaxy by integrating the best-fit single or double exponential $r$-band radial profile, and then calculate an equivalent 
disk scalelength $R_{\rm d, disk}$ as $R_{\rm e, disk}$/1.678.

To estimate the maximum rotation velocity, we construct absorption-line rotation curves by using the geometric parameters (i.e., center, ellipticity and PA) derived from our isophotal 
analysis.\ Inclination-corrected rotation velocity at a given radius of a galaxy is determined by fitting a sinusoidal function to the azimuthal distribution of the line-of-sight velocity 
measurements (retrieved from the MaNGA DAP products) of spaxels falling within an elliptical annulus.\ The maximum rotation velocity is then determined by fitting the rotation 
curve with a functional form given in \cite{Barrera-Ballesteros2018}.

\section{Results}\label{sec:results}

\subsection{Hubble types, bars and stellar masses of galaxies with different surface brightness profile types}

\begin{figure*}
\centering
\includegraphics[width=1\textwidth]{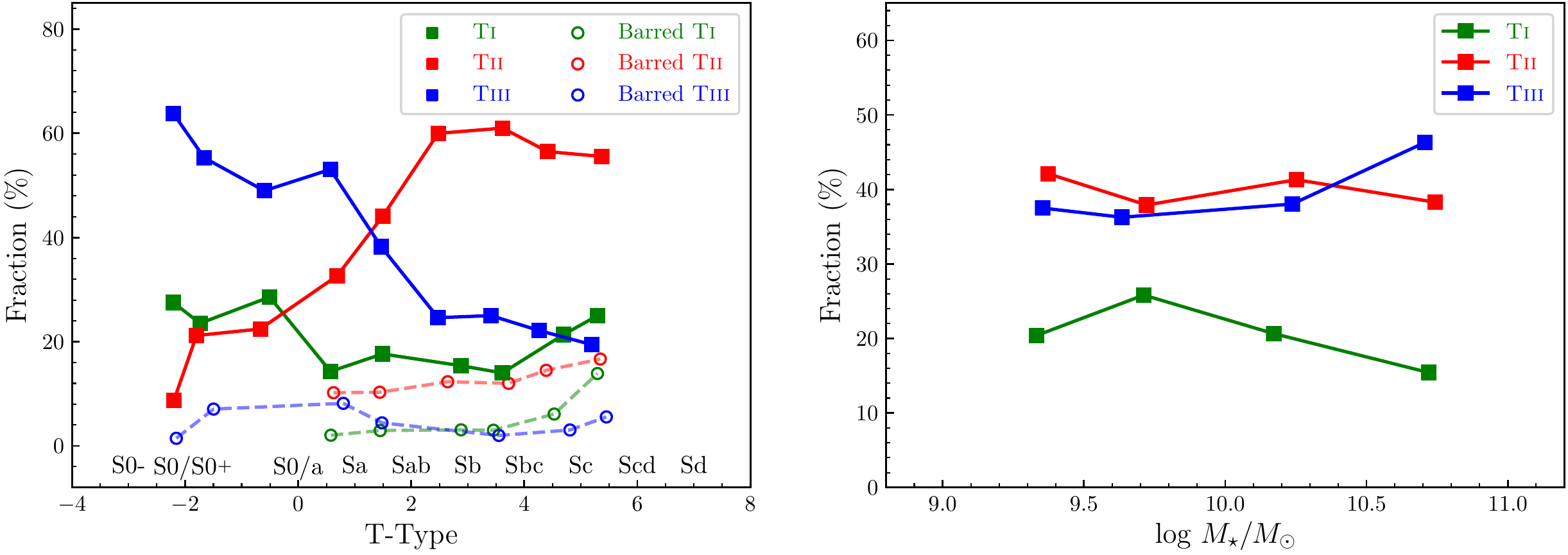}
\caption{{\it Left panel:} T-Type distributions of the three disk profile types in our sample.\ The correspondence between T-Types and Hubble morphological types is indicated at the 
bottom of the panel.\ Subsamples with probabilities of having bar signatures $>$ 50\% \citep{Fischer2019} are plotted as open symbols and dashed lines.\ 
{\it Right panel:} Stellar mass distributions of different disk profile types.\
The y-axis values are the fractions with respect to the total number of galaxies at given T-Type values ({\it left}) or stellar masses ({\it right}).}
\label{fig:disktype}
\end{figure*}

Among the 635 galaxies in our final sample, 152 (24\%) have {\sc Ti} surface brightness profiles, 264 (42\%) have {\sc Tii} profiles and 219 (34\%) have {\sc Tiii} profiles.\ 
The dominant fraction of double exponential profiles (76\%) is broadly in line with statistics from previous studies (e.g., $\sim$ 90\% in \citealt{Pohlen2006}, 88\% in 
\citealt{Gutierrez2011} and 84\% in \citealt{Marino2016}).\ Note that the higher fraction of double exponential profiles found in previous studies should 
be primarily attributed to an underrepresentation of early-type disk galaxies which have higher fraction of {\sc Ti} profiles than late-type galaxies \citep{Gutierrez2011}.\ 
Also recall that we have excluded triple exponential profiles from our final sample.\ Figure \ref{fig:disktype} shows the fraction of different disk profile types as a 
function of T-Type (left panel).\ The fraction of {\sc Tii} profiles increases toward later T-Types, reaching the maximum of $\sim$ 60\% at T-Type $\gtrsim$ 2 (i.e., Sab and later).\ 
In contrast to {\sc Tii} profiles, the fraction of {\sc Tiii} profiles increases toward earlier T-Types, peaking at $\sim$ 60\% at T-Type $\lesssim$ $-2$ (i.e.,\ S0 and earlier).\ The fraction 
of {\sc Ti} profiles ($\sim$ 20\%) at T-Type $<$ 0 is generally higher (by a factor of $\sim$ 2) than at later T-Types.\ All of these trends are in general agreement with previous studies 
\citep[e.g.,][]{Gutierrez2011}.

We use the MDLM-VAC P\_bar\_GZ2 parameter (the probability of having a bar signature) to select the most probable and (thus) strongly barred galaxies in our sample 
with P\_bar\_GZ2 $>$ 50\%, and plot their distributions separately in the left panel of Figure \ref{fig:disktype}.\ We can see that galaxies with strong bars account for a small 
fraction ($\lesssim$ $10-20$\%) of our samples at any given T-Types.\ {\sc Tii} galaxies appear to have a higher bar fraction than the other two profile types at T-Types 
$\gtrsim$ 0.5.\ Nevertheless, we point out that the actual fraction of barred galaxies, especially when taking into account of the weakly barred ones, can probably reach 
up to $\sim 40-70$\%, as suggested in a recent study by \cite{Erwin2018}.\ 

The fraction of different disk profile types generally has a weak stellar mass dependence (right panel of Figure \ref{fig:disktype}).\ The slight increase of the fraction of {\sc Tiii} 
profiles toward higher stellar masses probably reflects the well-known morphology-mass correlation \citep[e.g.,][]{Calvi2012}, whereby more massive galaxies are more likely 
to have earlier Hubble types (and thus a higher fraction of {\sc Tiii} profiles).\ In a similar vein, the slight decrease of the fraction of {\sc Ti} profiles toward higher stellar masses 
appears to be in line with a lower fraction of {\sc Ti} profiles at later Hubble types.\

\subsection{Mass-size relation for galaxies with different surface brightness profile types}\label{sec: massize}

\begin{figure*}
\centering
\includegraphics[width=1\textwidth]{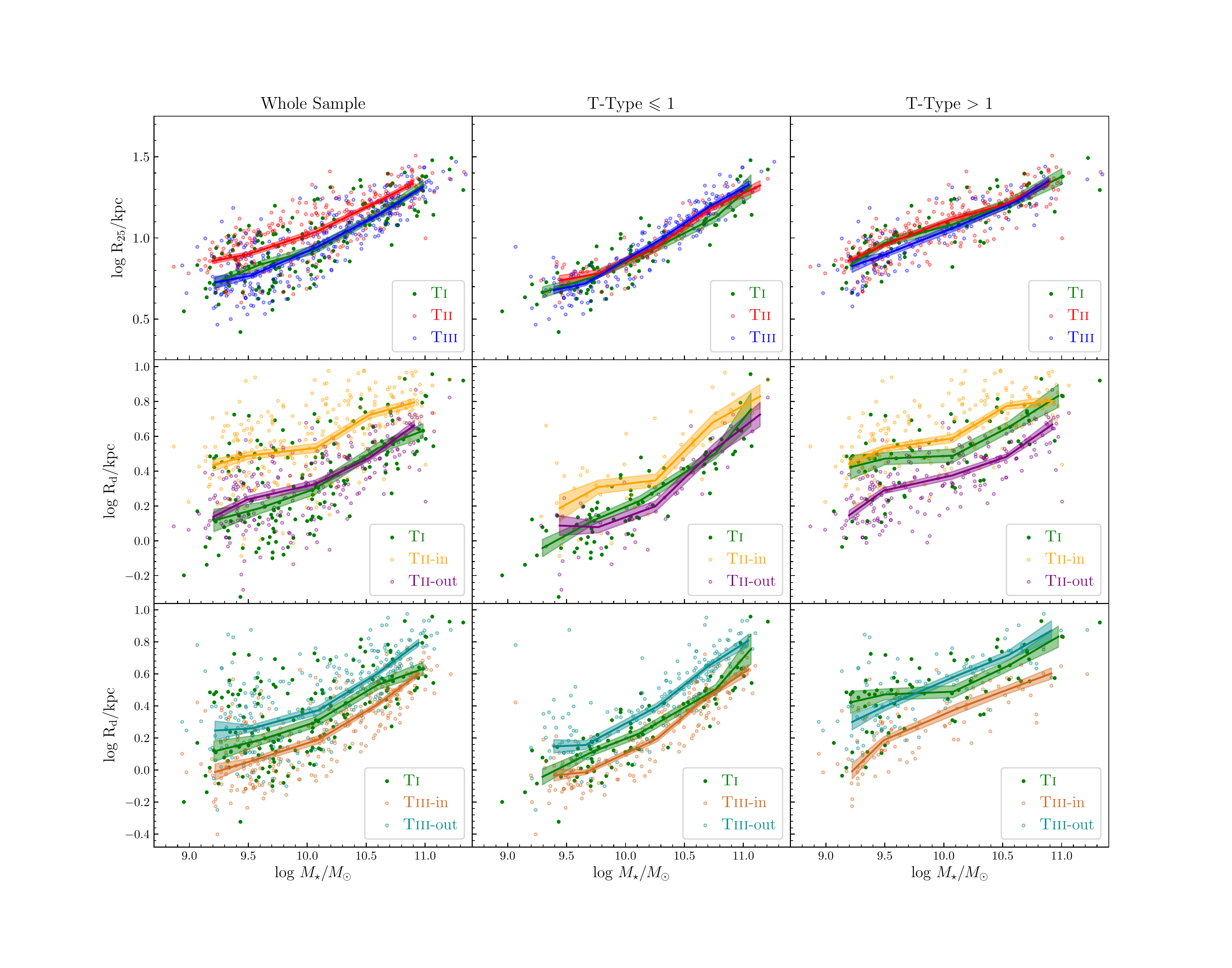}
\caption{
Stellar mass-size relations of galaxies with different disk profile types.\
{\it Top row:} stellar masses are plotted against the $g$-band 25 mag arcsec$^{-2}$ isophotal radius $\rtf$.\ 
{\it Middle row:} stellar masses are plotted against exponential disk scale lengths for the {\sc Ti} profiles (green), 
the inner disks (yellow) and outer disks (purple) of {\sc Tii} profiles.\
{\it Bottom row:} stellar masses are plotted against exponential disk scale lengths for the {\sc Ti} profiles (green), 
the inner disks (chocolate) and outer disks (darkcyan) of {\sc Tiii} profiles.\
Distributions for the whole sample, early-type (T-Type $\leqslant$ 1) and late-type (T-Type $>$ 1) subsamples are shown in the first, second and third column respectively.\
In each penal, medians of $\rtf$ (scale length $R_{\rm d}$) as a function of stellar masses are plotted as solid line.\ 
The shades represent the uncertainties of medians (RMS/$\sqrt{N}$), where $N$ is the total number of galaxies in a given bin.
}
\label{fig:massradius}
\end{figure*}

The distribution of our galaxies on the stellar mass-size planes is shown in Figure \ref{fig:massradius}.\
Instead of using the half-light radius $R_{e}$, which is a biased size measurement when comparing galaxies with different light concentrations (see below), we choose to 
use the the $g$-band 25 mag arcsec$^{-2}$ isophotal radius $\rtf$ to quantify the overall size of our galaxies.\ Note that the $g$-band 25 mag arcsec$^{-2}$ isophotes 
(on the AB system) reach $\sim$ $0.3-0.5$ mag (depending on galaxy types) fainter, and hence further in radius, than the familiar $B$-band 25 mag arcsec$^{-2}$ (on the 
Vega system) isophotes.\ We can see that galaxies with different disk profile types follow nearly the same $M_{\star}-\rtf$ relation, once the sample is divided into early-type 
(T-Type $\leqslant$ 1) and late-type (T-Type $>$ 1) disk galaxies.\ In addition, for given $\Ms$ and T-Types, galaxies of the three profile types have very similar median ($g-r$) colors 
at $\rtf$ (not shown here), with typical differences $\lesssim$ 0.05 mag, which suggests that they have nearly the same median stellar mass surface densities at $\rtf$, 
considering a general correlation between ($g-r$) and stellar mass-to-light ratios.\ It is worth mentioning that previous studies found similar surface mass densities at the 
break radii of different profile types \citep{Bakos2008,Herrmann2016}. \ The similar mass-$\rtf$ relations for galaxies with different profile types make the hypothesis 
that stellar radial migration alone may transform {\sc Tii} to {\sc Ti} and finally to {\sc Tiii} profiles highly unlikely.\ 

To understand the formation processes of disk breaks, it is helpful to compare the exponential disk scale lengths $\rd$ of {\sc Ti} profiles to that of the inner- and outer-exponential  
disk $\rd$ of broken profiles.\ Such comparisons are shown in the middle and bottom rows of Figure \ref{fig:massradius}.\ By dividing the sample into early-type and late-type 
galaxies, it is obvious that {\sc Tii} galaxies (middle row) generally have larger inner-disk median $\rd$ than {\sc Ti} galaxies, whereas {\sc Tiii} galaxies (bottom row) 
generally have smaller inner-disk median $\rd$ than {\sc Ti} galaxies, irrespective of the morphological types.\ On the other hand, the outer-disk median $\rd$ of early-type (late-type) 
{\sc Tii} galaxies is generally similar to (smaller than) the $\rd$ of {\sc Ti} galaxies, while the outer-disk median $\rd$ of {\sc Tiii} galaxies is mostly larger than that of {\sc Ti} galaxies.\
Above all, neither the inner-disk $\rd$ nor the outer-disk $\rd$ is an appropriate size parameter to use when comparing galaxies with different disk profile types and morphological types.\ 
We also note that the scatter of the mass-$\rd$ relations is generally larger than that of the mass-$\rtf$ relations.\

\subsection{Radial locations of surface brightness profile breaks}\label{sec:rbreak}

\begin{figure*}
\centering
\includegraphics[width=1\textwidth]{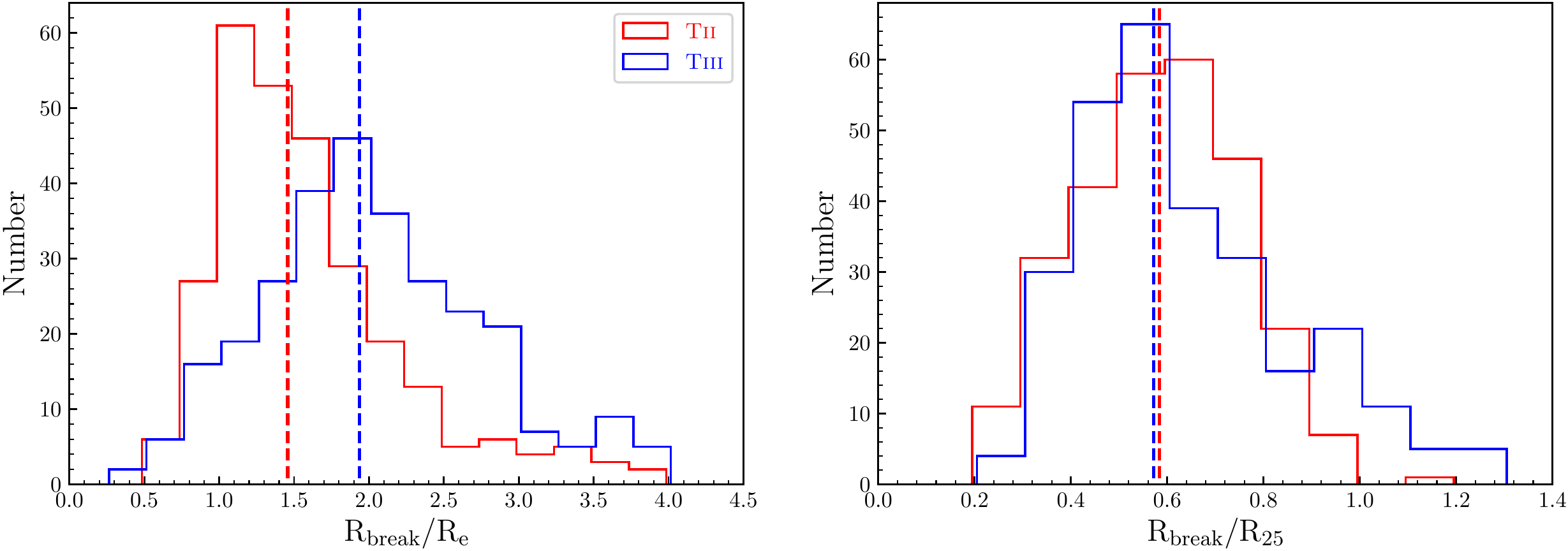}
\caption{
Distribution of disk break radius $R_{\rm break}$ for {\sc Tii} (red) and {\sc Tiii} galaxies (blue).\ 
$R_{\rm break}$ is normalized by the $r$-band half-light radius $\re$ in the {\it left panel} and by the $g$-band $\rtf$ in the {\it right panel}.\
The red and blue vertical dashed lines in each panel mark the median for {\sc Tii} and {\sc Tiii} galaxies respectively.
}
\label{fig:hisrbreak}
\end{figure*}

The surface brightness profile breaks of our {\sc Tii} and {\sc Tiii} galaxies are located at a range of galactocentric radii from 1.3 kpc to 23.3 kpc.\ We present the distribution of  
the break radii $\rbreak$ normalized by $\re$ and $\rtf$ respectively in the left and right panels of Figure \ref{fig:hisrbreak}.\ The $\rbreak/\re$ distribution of {\sc Tii} galaxies 
peaks at significantly smaller values than that of {\sc Tiii} galaxies, with a median $\rbreak/\re$ of 1.47 for the {\sc Tii} galaxies and 1.92 for the {\sc Tiii} galaxies.\ In contrast to 
$\rbreak/\re$, $\rbreak/\rtf$ of {\sc Tii} and {\sc Tiii} galaxies have very similar distributions, with virtually the same median values (0.58$\pm$0.16 and 0.57$\pm$0.23).

\subsection{Radial profiles of surface brightness, spectral features and $\Sigma_{\star}$} \label{sec:profiles}

\begin{figure*}
\centering
\includegraphics[width=1\textwidth]{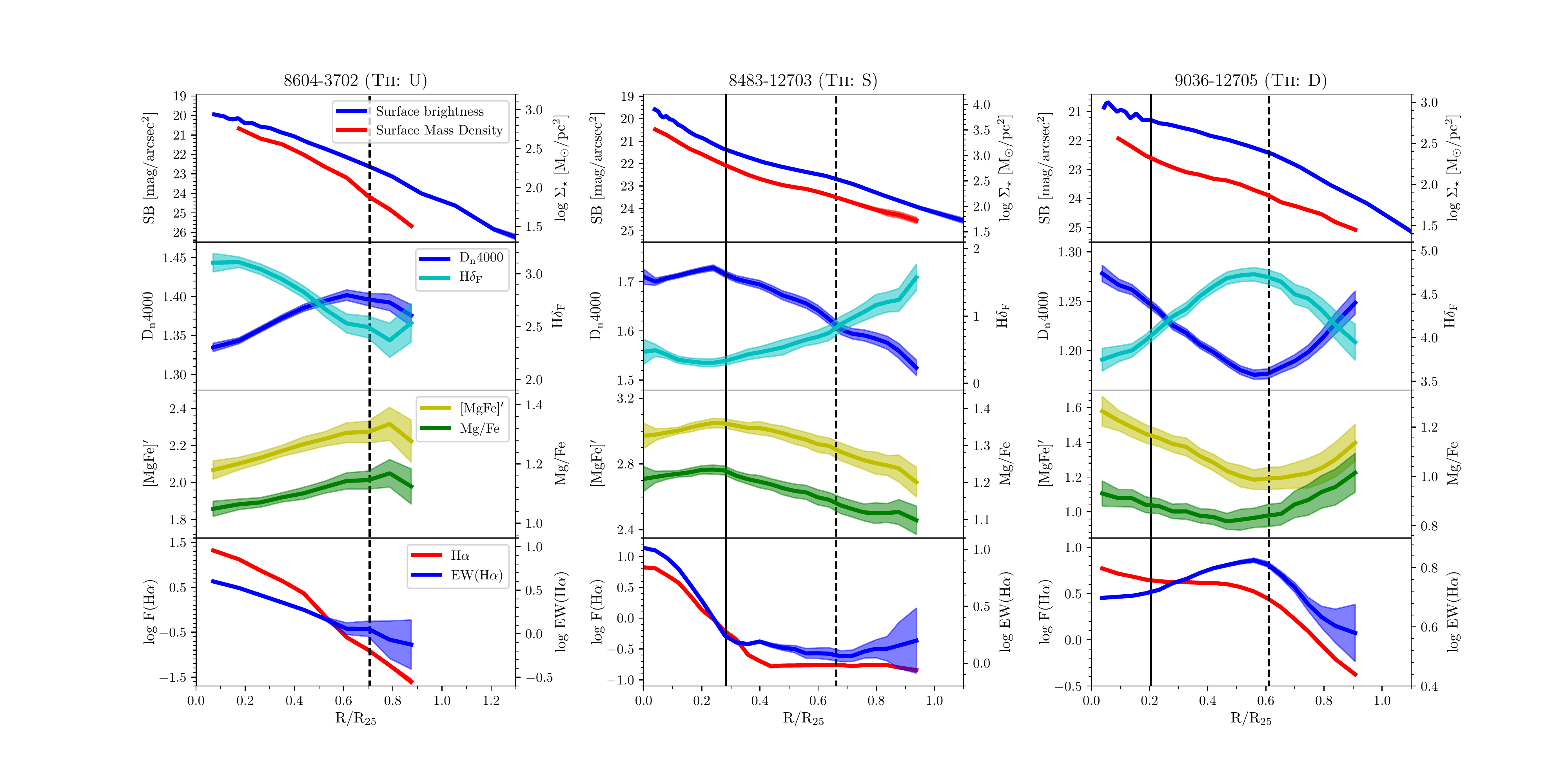}
\caption{
Examples of radial profiles of various quantities for the subsample with {\sc Tii} surface brightness profiles.\
The left column corresponds to a galaxy with up-bending $\hdeltaf$ profiles ({\sc Tii}: U), the middle column corresponds to a galaxy with simple linear declining $\hdeltaf$ profile ({\sc Tii}: S), and the right column corresponds to a galaxy with an down-bending $\hdeltaf$ profile ({\sc Tii}: D).\ For each galaxy, the first row shows the surface brightness and 
stellar mass density $\Sigma_{\star}$ profiles, the second row shows the radial profiles of $\dindex$ and $\hdeltaf$ indices, the third row shows the radial profiles of [MgFe]$^{\prime}$ 
and Mg/Fe indices, and the bottom row shows the radial profiles of H$\alpha$ flux densities and EW(H$\alpha$).\ In each panel, the solid vertical line marks the bulge-to-disc 
transition radius (Section \ref{sec:sbpfit}), which sets the inner bound of the radial range used for our disc surface brightness profile fitting. The vertical dashed line in each panel 
marks the radius of surface brightness profile break. Examples of {\sc Ti} and {\sc Tiii} galaxies are shown in the appendix (Figures \ref{fig:examplet1} and \ref{fig:examplet3}).
}
\label{fig:examplet2}
\end{figure*}

\begin{deluxetable*}{lccccc}
\tabletypesize{\small}
\tablecolumns{6}
\setlength{\columnsep}{0.002pt}
\tablewidth{0pt}
\tablecaption{$\hdeltaf$ radial profile shapes}
\tablehead{
\colhead{Sample} 
& \multicolumn{5}{c}{Profile shapes} \\
\cline{2-6}
\colhead{}
&\colhead{Up-bending}
&\colhead{Single}
&\colhead{Down-bending-D}
&\colhead{Down-bending-F}
&\colhead{Down-bending-$\Lambda$} \\
\noalign{\vskip -4.5mm}
}

\startdata
Early-type {\sc Ti} & 16(20\%) & 39(49\%) & 4(5\%) & 9(11\%) & 11(14\%) \\
Early-type {\sc Tii} & 8(17\%) & 21(44\%) & 1(2\%) & 8(17\%) & 11(22\%) \\
Early-type {\sc Tiii} & 20(15\%) & 73(55\%) & 5(4\%) & 25(19\%) & 10(8\%) \\
Late-type {\sc Ti} & 6(8\%) & 24(33\%) & 9(12\%) & 24(33\%) & 10(14\%) \\
Late-type {\sc Tii} & 23(12\%) & 38(19\%) & 22(11\%) & 56(28\%) & 60(30\%) \\
Late-type {\sc Tiii} & 13(22\%) & 12(20\%) & 6(10\%) & 21(35\%) & 8(13\%) \\
\enddata
\tablecomments{See Section \ref{sec:classspectr} or Figure \ref{fig:hdeltafprofileclass} for the definition of different profile shapes, and  
Section \ref{sec:profiles} for a discussion on the overall trend.}
\label{tab:hdeltaclass}
\end{deluxetable*}

\begin{figure*}
\centering
\includegraphics[width=1\textwidth]{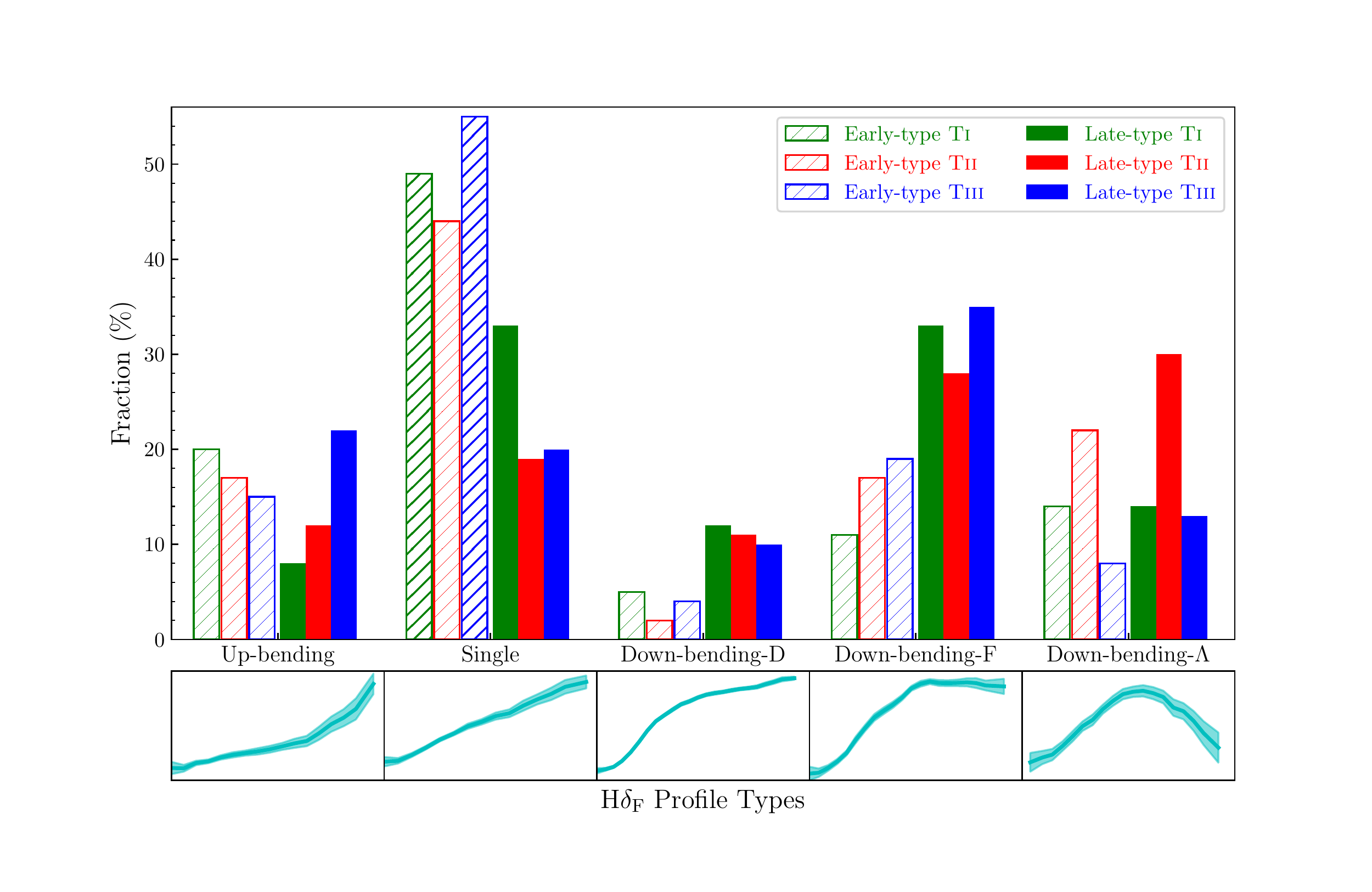}
\caption{
Distribution of the overall shapes of $\hdeltaf$ radial profiles for galaxies with T-Type $\leqslant$ 1 (early-type: hatched bars) and T-Type $>$ 1 (late-type: filled bars).\
The y-axis fractions are with respect to the total number of galaxies with given T-Type and surface brightness profile type.\
``Up-bending'' means the radial gradients become shallower beyond the break radii, ``Single'' means a single linear declining with radius, ``Down-bending-D'' means 
the radial gradients become shallower but do not change direction beyond the break radii, ``Down-bending-F'' means the radial gradients become flat beyond the 
break radii, and ``Down-bending-$\Lambda$'' means approximately $\Lambda$-shaped radial profiles.\ Illustrative examples for the five profile shapes are shown 
below the histograms.\ 
}
\label{fig:hdeltafprofileclass}
\end{figure*}

Representative examples of radial profiles of the $r$-band surface brightness, $\Sigma_{\star}$ and our interested spectral features for the {\sc Ti}, {\sc Tii} and {\sc Tiii} galaxies 
are shown in Figures \ref{fig:examplet1}, \ref{fig:examplet2} and \ref{fig:examplet3} respectively.\ The breaks of surface brightness profiles and other profiles explored here, 
whenever they exist, are in reasonable agreement in their radial locations.\ As described in Section \ref{sec:classspectr}, we make a classification of the basic shapes of $\hdeltaf$ 
radial profiles.\ The results of the classification are illustrated in Figure \ref{fig:hdeltafprofileclass} and are also given in Table \ref{tab:hdeltaclass}.\ Generally speaking, 
down-bending $\hdeltaf$ radial profiles correspond to up-bending stellar age radial profiles, and down-bending $\Lambda$-shape $\hdeltaf$ radial profiles correspond 
to V-shape (or the commonly denoted U-shape) stellar age radial profile.\

Galaxies exhibit a diversity of radial profiles of the age/metallicity-sensitive spectral features and $\Sigma_{\star}$, irrespective of their surface brightness profile types.\ 
Similar findings have been reported previously \citep[e.g.,][]{Roediger2012, Ruiz-Lara2016}, but with an order of magnitude larger and much more homogeneous sample 
than previous studies, we can now obtain relatively robust statistics on the frequency of different stellar age (as traced by $\hdeltaf$) profile types for the first time.\ As shown 
in Figure \ref{fig:hdeltafprofileclass}, the distribution of $\hdeltaf$ profile shapes have an obvious dependence on Hubble types, in the sense that the single linear $\hdeltaf$ 
profile shape is the most common one among early-type disk galaxies (T-Type $\leqslant$ 1), irrespective of surface brightness profile types, whereas the 
down-bending $\hdeltaf$ profiles, including down-bending-F and down-bending-$\Lambda$, are the most common ones among late-type disk galaxies  (T-Type $>$ 1), again 
irrespective of surface brightness profile types.\ In addition, the familiar U-shape stellar age profiles, as represented by the $\Lambda$-shape $\hdeltaf$ profiles here, are the 
most common ones only in late-type {\sc Tii} galaxies.\ 

\subsection{Radial gradient slopes $\alpha$ of inner versus outer disks}\label{sec:slopes}

\begin{figure*}
\centering
\includegraphics[width=1\textwidth]{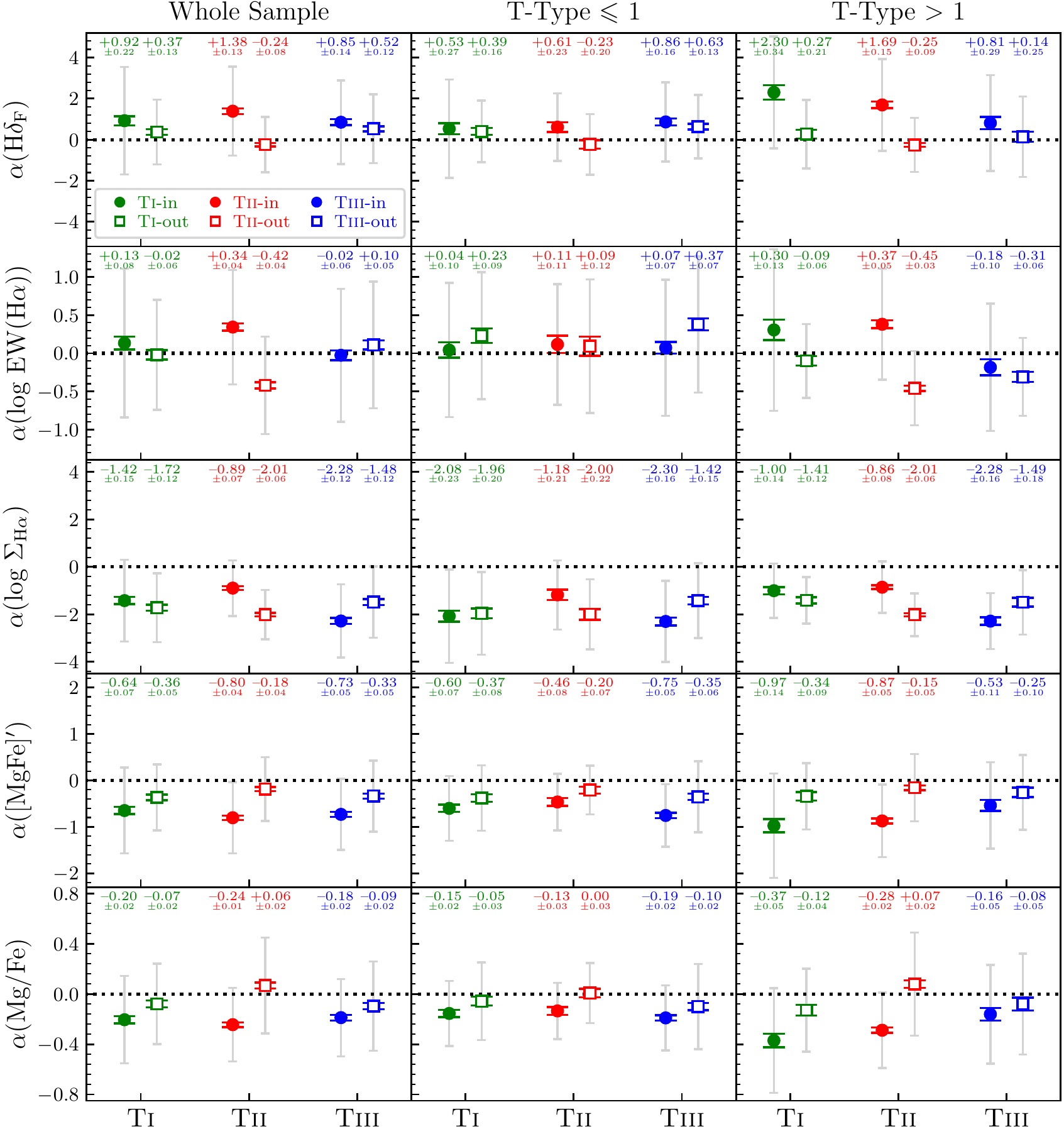}
\caption{
Median radial gradient slopes $\alpha$ (per $\rtf$) of different spectral features or indices for the whole sample (left), early-type subsample (middle) and late-type subsample (right).\  
Each row is for one spectral index or feature, as indicated in the y-axis title.\ $\alpha$ for the inner disks is plotted as filled symbols while $\alpha$ for the outer disks as open symbols.\
The grey error bars are the RMS scatters of the distributions of individual galaxies, while the color error bars represent the statistical uncertainties of medians (RMS/$\sqrt{N}$), where 
$N$ is the total number of galaxies in a given subsample.\ The median values and their uncertainties are also given at the top of each panel.}
\label{fig:gradinout}
\end{figure*}

Generally speaking, stellar radial migration induced by either resonant scattering of non-axisymmetric structures (e.g., bars, spiral arms) or environmental influences 
(e.g., satellite accretion) tends to weaken stellar population gradients (see references in the Introduction section).\ In order to test whether different surface brightness 
profile types can be qualitatively explained by different efficiencies of stellar radial migration, Figure \ref{fig:gradinout} compares radial gradient slopes $\alpha$ of age-, 
metallicity- and SFR-sensitive spectral features for galaxies with different surface brightness profile types.\ Both the scatter of various $\alpha$ distributions (grey error bars) 
and statistical uncertainties of the medians (green, red and blue error bars) are indicated in Figure \ref{fig:gradinout}.

We first discuss the median $\alpha$ trend for the age-sensitive features.\ The inner disks of {\sc Tii} galaxies (regardless of the morphological types) have positive 
median $\alpha$($\hdeltaf$) that are opposite to the outer disks, which corresponds to the familiar U-shape or V-shape stellar age profiles.\ $\alpha$(log EW(H$\alpha$)) of {\sc Tii} galaxies follows a similar inner-outer contrast as $\alpha$($\hdeltaf$), with an apparent exception of the early-type ones (T-Type $\leqslant$ 1) which have nearly the same inner and outer median $\alpha$(log EW(H$\alpha$)) within uncertainties.\ {\sc Tii} galaxies have the steepest (positive) inner-disk median $\alpha$($\hdeltaf$) and $\alpha$(log EW(H$\alpha$)) among the three profile types for the whole sample, which seems consistent with the idea that {\sc Tii} profiles are the least influenced by stellar migration effect, but it is not the case once splitting the whole sample into early and late morphological types.\ For the early-type subsamples, the three break types appear to have similar inner-disk median $\alpha$($\hdeltaf$) and $\alpha$(log EW(H$\alpha$)), and their outer-disk median gradients are also very close to each other when considering the statistical uncertainties.\ For the late-type subsamples, the {\sc Ti} and {\sc Tii} galaxies have similar positive inner-disk median $\alpha$($\hdeltaf$) and $\alpha$(log EW(H$\alpha$)) that are significantly steeper than that of the {\sc Tiii} galaxies.\

Regarding the {\it in situ} star formation distributions as traced by $\Sigma_{\rm H\alpha}$ profiles, the outer disks of {\sc Tii} ({\sc Tiii}) galaxies have steeper (shallower) median 
$\alpha$(log $\Sigma_{\rm H\alpha}$) than the inner disks, which is an unambiguous evidence that abrupt changes in star formation intensities from the inner disks to outer disks 
contribute to the formation of both down-bending and up-bending breaks, irrespective of morphological types.\ This also rules out the possibility that a superposition of thin and 
thick disks with different scale lengths is a dominant mechanism for shaping {\sc Tiii} profiles in our sample.\ Moreover, late-type {\sc Tiii} galaxies have significantly more negative 
inner-disk median $\alpha$(log $\Sigma_{\rm H\alpha}$) than late-type {\sc Ti} and {\sc Tii} galaxies, suggesting that star formation in {\sc Tiii} galaxies is usually much more concentrated 
toward smaller radii.\ For the early-type subsample, the {\sc Tii} and {\sc Tiii} galaxies have about the same inner and outer $\alpha$(log $\Sigma_{\rm H\alpha}$) as their late-type 
counterparts, but the {\sc Ti} galaxies have significantly more negative inner and outer median $\alpha$(log $\Sigma_{\rm H\alpha}$) than their late-type counterparts.\

Regarding the metallicities, late-type {\sc Ti} and {\sc Tii} galaxies have significantly steeper inner-disk median $\alpha$([MgFe]$^{\prime}$) than late-type {\sc Tiii} galaxies, whereas the reverse is true for early-type galaxies.\ {\sc Ti} and {\sc Tii} galaxies have comparable inner-disk median $\alpha$([MgFe]$^{\prime}$) within uncertainties.\ For the outer disks, {\sc Tii} galaxies have slightly shallower $\alpha$([MgFe]$^{\prime}$) than {\sc Ti} and {\sc Tiii} galaxies.\ {\sc Ti} and {\sc Tiii} galaxies have comparable outer-disk median $\alpha$([MgFe]$^{\prime}$) within uncertainties.\ It is worth noting that the outer-disk median $\alpha$([MgFe]$^{\prime}$) is always shallower than the inner-disk $\alpha$([MgFe]$^{\prime}$) for any given profile type, which indicates either a non-negligible effect of stellar radial migration or satellite accretion for {\sc Ti} and {\sc Tii} galaxies, because otherwise we would expect the outer-disk metallicity gradients to be similar to or steeper than the inner-disk gradients \citep[e.g.,][]{Sanchez-Blazquez2009}.\ For {\sc Tiii} galaxies, however, the relatively steeper inner-disk median $\alpha$([MgFe]$^{\prime}$) may be primarily ascribed to a steeper radial decline of star formation efficiency at the inner disks.\ 
All of the above trends for $\alpha$([MgFe]$^{\prime}$) generally apply to $\alpha$(Mg/Fe).\

Based on 214 spiral galaxies from the CALIFA survey, \cite{Ruiz-Lara2017} found that {\sc Tii} galaxies show the steepest inner-disk radial gradients of stellar ages and metallicities, {\sc Tiii} galaxies show the shallowest, and {\sc Ti} galaxies have radial gradients in between that of {\sc Tii} and {\sc Tiii} galaxies.\ With nearly 3 times larger sample size and a much more uniform coverage of stellar masses and morphological types than the \cite{Ruiz-Lara2017} study, we find a more complicated dependence of stellar population gradients on disk profile types.\ In particular, late-type {\sc Tii} galaxies do not have significantly steeper median gradient slopes than {\sc Ti} galaxies, and early-type {\sc Tiii} galaxies have steeper, instead of shallower, inner-disk median metallicity gradient slopes than the other two profile types.\ Stellar radial migration is expected to gradually weaken stellar population gradients.\ Therefore, our finding negates the simple hypothesis that stellar radial migration alone can transform profile types from {\sc Tii} to {\sc Ti} and then to {\sc Tiii}, in line with our conclusion drawn in Section \ref{sec: massize}.

\subsection{Radial profile break strengths}

\begin{figure*}
\centering
\includegraphics[width=1\textwidth]{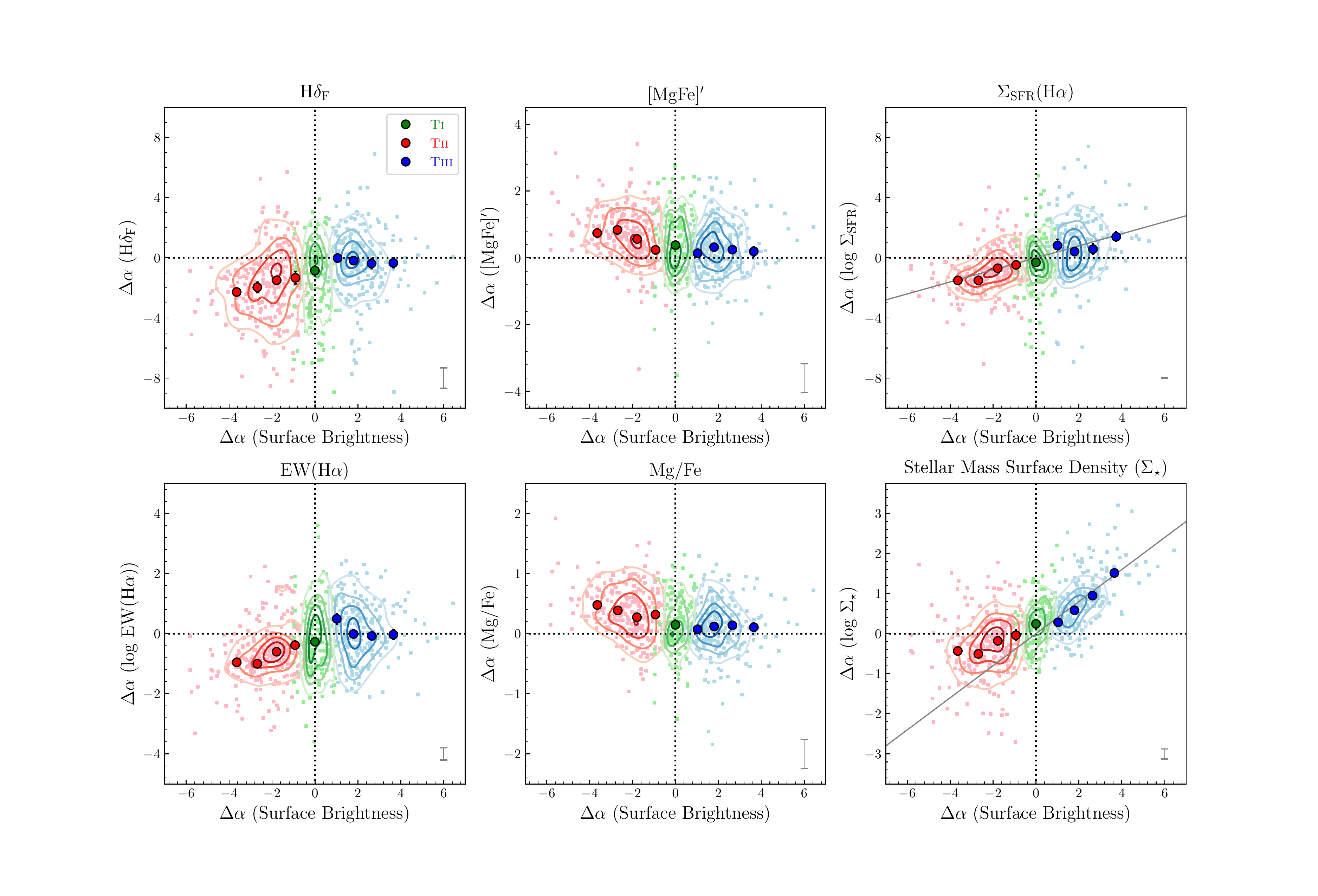}
\caption{
Break strengths of the radial profiles of $\hdeltaf$, [MgFe]$^{\prime}$, Mg/Fe, $\Sigma_{\rm SFR}$(H$\alpha$), EW(H$\alpha$) and $\Sigma_{\star}$ are plotted against that 
of the surface brightness profiles.\ In each panel, the small squares represent individual galaxies (light green: {\sc Ti}; light red: {\sc Tii}; light blue: {\sc Tiii}), while the big circles 
represent the medians as a function of surface brightness profile break strengths for different profile types (green: {\sc Ti}; red: {\sc Tii}; blue: {\sc Tiii}).\ Also overplotted in each 
panel are the number density contours for different profile types.\ The contours are drawn at intervals of 20\% the peak number densities.\ The break strengths $\Delta$$\alpha$ 
are defined as the outer minus inner disk gradient slopes (per $\rtf$), where the outer and inner disks are demarcated by the break radii.\ A typical error bar for the break strength 
measurements is shown at the bottom right corner of each panel.\ Note that, besides the {\sc Tii} and {\sc Tiii} galaxies, {\sc Ti} galaxies which have down-bending or up-bending 
$\hdeltaf$ profiles are also plotted, and in these cases, the $\hdeltaf$ profile break radii are used for calculating break strengths of the other profiles.\ In each of the two rightmost 
panels, a line with a slope of 0.4 and a y-intercept of 0 is over-plotted to represent the relation expected if $\Sigma_{\rm SFR}$(H$\alpha$) or $\Sigma_{\star}$ is proportional to 
the stellar light intensity.
}
\label{fig:bkstrength_comp1}
\end{figure*}

\begin{figure*}
\centering
\includegraphics[width=0.9\textwidth]{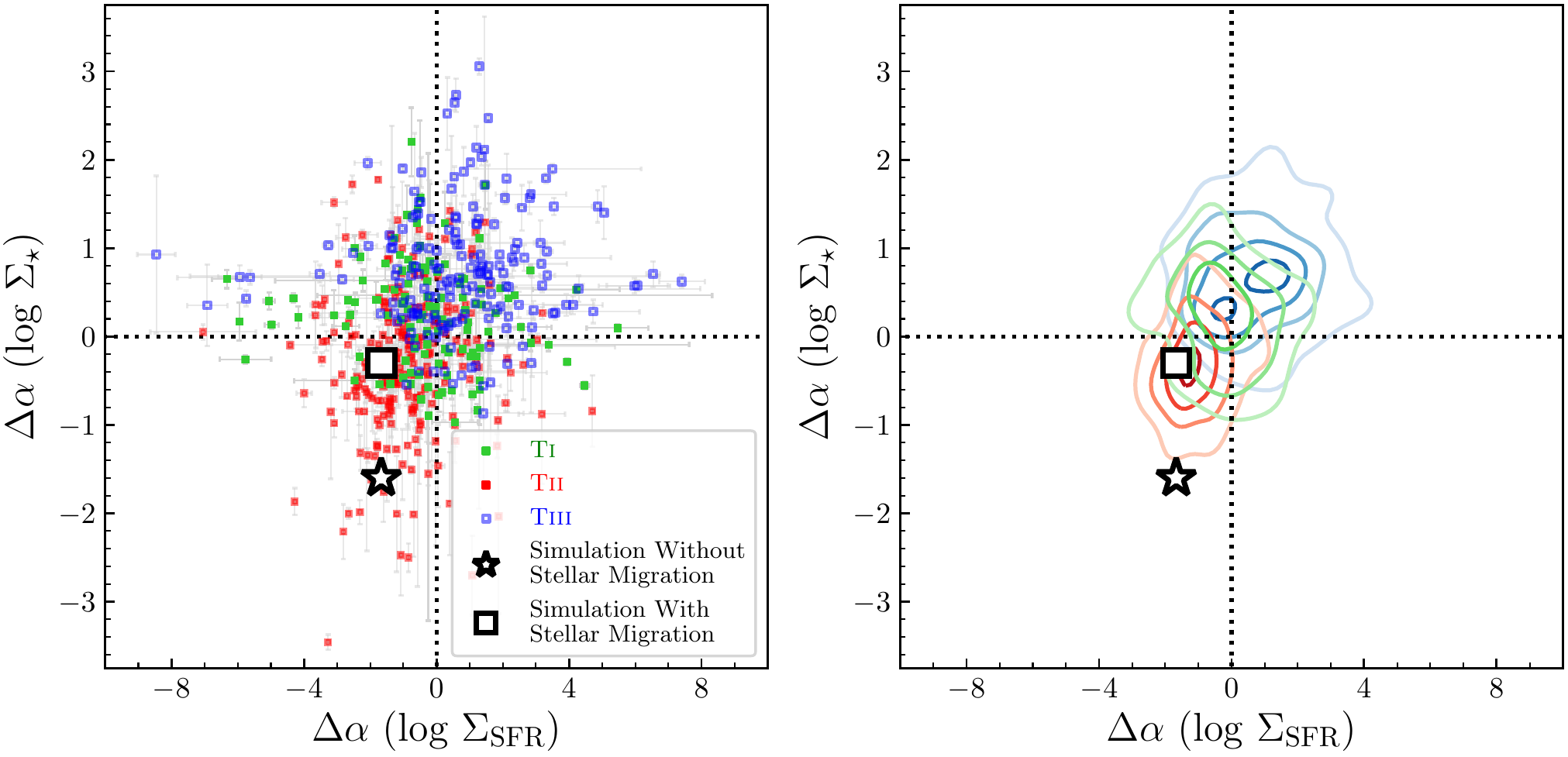}
\caption{
Break strengths of $\Sigma_{\rm SFR}$ versus $\Sigma_{\star}$ profiles.\
The left panel shows the scatter plot, while the right panel shows the corresponding number density contours with contour intervals of 20\% the peak number densities.\ 
In both panels, {\sc Ti}, {\sc Tii} and {\sc Tiii} subsamples are respectively shown as green, red and blue colors.\ The white filled square and white filled star symbols, respectively, 
mark the predictions of the \cite{Sanchez-Blazquez2009} galaxy disk formation simulations with and without stellar radial migration.\ 
}
\label{fig:dslope_sfr_mass}
\end{figure*}

Different formation scenarios of disk breaks generally predict different behaviors of stellar population radial variations across the disks.\ In this section, we use the outer 
minus inner radial gradient slope differences $\Delta$$\alpha$ to quantify the break strengths of our galaxies, and explore the relationship between the radial profile break 
strengths of spectral features, $\Sigma_{\rm SFR}$, $\Sigma_{\star}$ and surface brightness.\ Note that we use the break radii measured based on the surface brightness 
profiles for calculating the break strengths of other profiles.\ {\sc Ti} galaxies with down-bending or up-bending $\hdeltaf$ profiles are also included in the following comparisons, 
and for these galaxies, the $\hdeltaf$ profile break radii are used for calculating break strengths.\ For the sake of brevity, we do not distinguish between early-type and late-type 
galaxies in this section, but point out that the differences (see below) between early-type galaxies with different profile types are generally significantly smaller than that between late-type galaxies.\

\subsubsection{Break strengths of surface brightness profiles versus other profiles}

Figure \ref{fig:bkstrength_comp1} compares the break strengths of surface brightness profiles to that of the spectral features, $\Sigma_{\rm SFR}$ and $\Sigma_{\star}$ profiles.\ 
{\sc Tii} galaxies have mostly negative $\Delta$$\alpha$($\hdeltaf$), positive $\Delta$$\alpha$([MgFe]$^{\prime}$), positive $\Delta$$\alpha$(Mg/Fe), negative 
$\Delta$$\alpha$(log $\Sigma_{\rm SFR}$) and negative $\Delta$$\alpha$(log EW(H$\alpha$)).\ Moreover, {\sc Tii} galaxies with stronger surface brightness profile breaks (i.e.,\ more 
negative $\Delta$$\alpha$) also have stronger median break strengths in the other explored radial profiles.\ In particular, the correspondence between the break strengths of 
$\Sigma_{\rm SFR}$ and surface brightness profiles for the whole sample is largely consistent with that expected if SFR is proportional to the stellar light intensity (the top 
right panel of Figure \ref{fig:bkstrength_comp1}).\ However, the $\Delta$$\alpha$(log $\Sigma_{\star}$) distribution of {\sc Tii} galaxies is skewed toward weaker break strengths 
compared to that would be expected if $\Sigma_{\star}$ is proportional to stellar light intensity (the bottom right panel of Figure \ref{fig:bkstrength_comp1}).\ 

Our finding for a systematically weaker $\Sigma_{\star}$ break strengths than that of the surface brightness profiles is in the similar sense to previous studies which, based on 
broadband optical photometry, found a near absence of breaks in the co-added $\Sigma_{\star}$ profiles of {\sc Tii} galaxies \citep[e.g.,][]{Bakos2008}, but here we emphasize that 
galaxies with {\sc Tii} surface brightness profiles generally have {\sc Tii} $\Sigma_{\star}$ profiles.\ As we will show in Section \ref{sec:brk_sfrmass}, stellar radial migration may play 
an important role in weakening but not completely erasing the {\sc Tii} $\Sigma_{\star}$ breaks.\ We mention in passing that dwarf galaxies with {\sc Tii} surface brightness profiles 
also generally have {\sc Tii} $\Sigma_{\star}$ profiles \citep{Zhang2012, Herrmann2016}.\

Contrary to the {\sc Tii} galaxies, {\sc Tiii} galaxies generally have weak or no correspondence between $\Delta$$\alpha$ of surface brightness profiles and the explored spectral 
feature profiles.\ Nevertheless, there is a significant positive correlation between the break strengths of surface brightness and $\Sigma_{\star}$ profiles of {\sc Tiii} galaxies, and the sense of this correlation is largely in line with that expected if $\Sigma_{\star}$ is proportional to stellar light intensity (i.e.\ the mass-to-light ratios stay more or 
less constant with radius), which is clearly illustrated in the bottom right panel of Figure \ref{fig:bkstrength_comp1}.\ {\sc Ti} galaxies have close to zero median break strengths 
for all of the radial profiles explored here.\

\subsubsection{Break strengths of $\Sigma_{\rm SFR}$ versus $\Sigma_{\star}$ profiles}\label{sec:brk_sfrmass}
Disk profile breaks that were produced by an abrupt change in star formation intensities may be expected to show a good correspondence between the break strengths 
of $\Sigma_{\rm SFR}$ and $\Sigma_{\star}$ profiles, whereas those formed through or significantly affected by stellar redistribution or accretion may not be expected to 
have such a correspondence.\ Figure \ref{fig:dslope_sfr_mass} shows the $\Delta$$\alpha$(log $\Sigma_{\rm SFR}$)$-$$\Delta$$\alpha$(log $\Sigma_{\star}$) distributions for different 
surface brightness profile types.\ The SFR surface densities are estimated from extinction-corrected H$\alpha$ flux densities.\ We compare our observations with predictions 
of a fully cosmological hydrodynamical simulation of galaxy disk formation by \cite{Sanchez-Blazquez2009}.\ 

According to the simulation of \cite{Sanchez-Blazquez2009}, the down-bending break in surface brightness profile owes its origin primarily to an abrupt decrease in gas volume 
density and thus in star formation efficiencies at large galactocentric distances.\ The stellar radial migration, which is induced either by non-axisymmetric structures in 
galaxies (e.g., bars and spiral arms) or environmental disturbances, tends to weaken the break strengths in stellar mass surface density profiles over time.\ 

The simulation results shown in Figure \ref{fig:dslope_sfr_mass} correspond to the predicted radial profiles at redshift $z = 0$ (Figures 4 and 16 in \citealt{Sanchez-Blazquez2009}).\
The result from the simulation with stellar migration falls near the peak of the observed $\Delta$$\alpha$(log $\Sigma_{\rm SFR}$)$-$$\Delta$$\alpha$(log $\Sigma_{\star}$) distribution of {\sc Tii} 
galaxies, whereas the simulation without stellar migration fails to match observations of any disk profile types.\ According to the \cite{Sanchez-Blazquez2009} simulation involving 
stellar radial migration, $\sim$ 60\% of the stars presently located beyond the break radius have migrated from the inner disk.\ We note that the \cite{Sanchez-Blazquez2009} simulations 
are limited to a low-spin Milky Way-like disk galaxy.\ It remains to be seen whether similar simulations covering a large parameter space (e.g., spin, mass and environment) can match 
the observations for all of the disk profile types.\

\subsection{Relevance of the spin parameter to disk profile types}

\begin{figure*}
\centering
\includegraphics[width=1\textwidth]{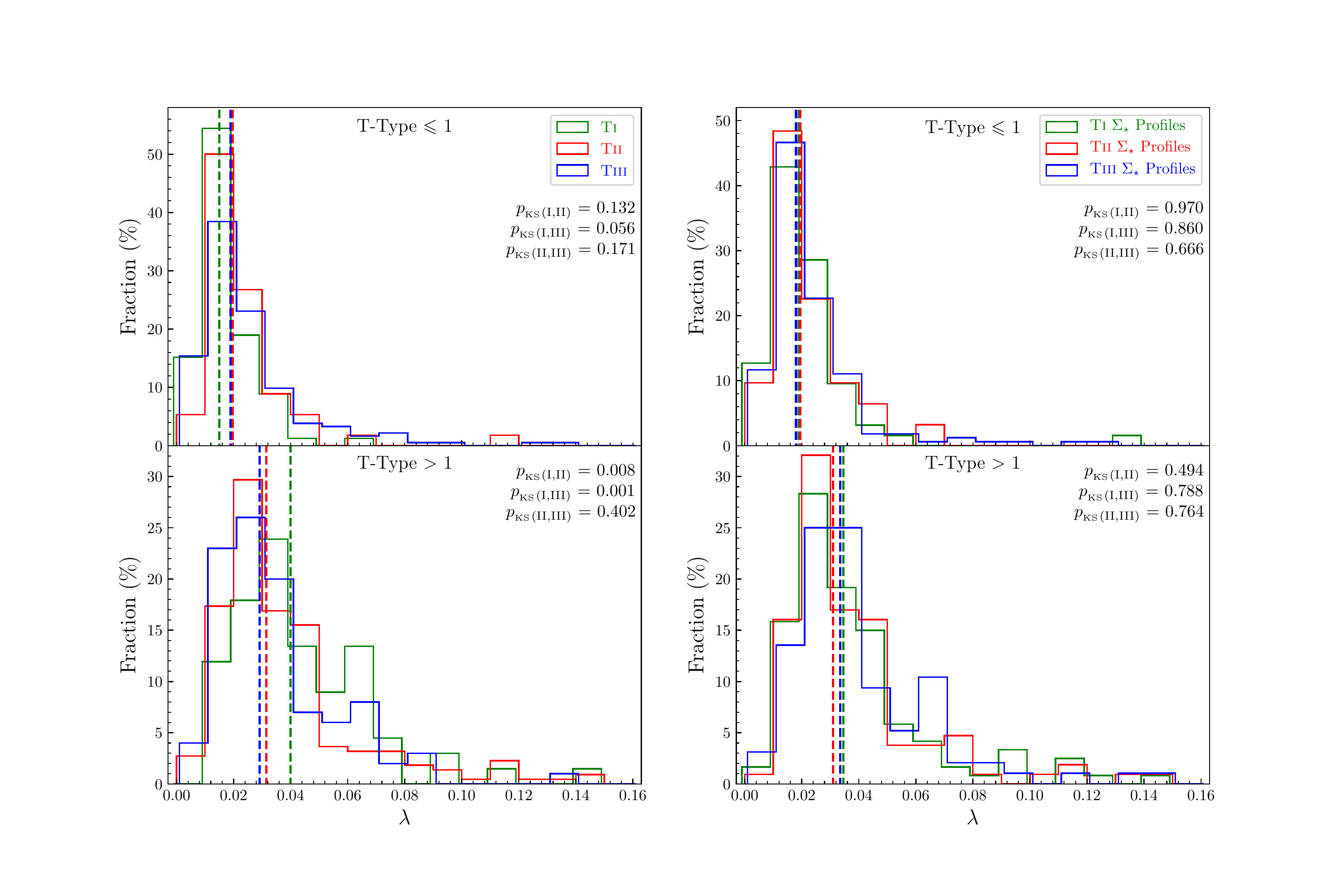}
\caption{
Distribution of spin parameters for galaxies with different surface brightness profile types (left column) and stellar mass surface density profiles (right column).\ 
Early-type galaxies (T-Type $\leqslant$ 1) are shown in the top panels while late-type galaxies (T-Type $>$ 1) are shown in the bottom panels.\ The $p$-values returned 
by the Kolmogorov-Smirnov test for subsamples of different profile types are listed in the figure.\ The vertical dashed lines mark the median $\lambda$ values of 
different subsamples.\ There is generally no significant difference (i.e., $p$ $>$ 0.05) between different profile types at given T-Type ranges, except that late-type 
{\sc Ti} galaxies appear to have significantly larger $\lambda$ than late-type {\sc Tii} and {\sc Tiii} galaxies.\ 
}
\label{fig:spinhist}
\end{figure*}

\begin{figure}
\centering
\includegraphics[width=0.46\textwidth]{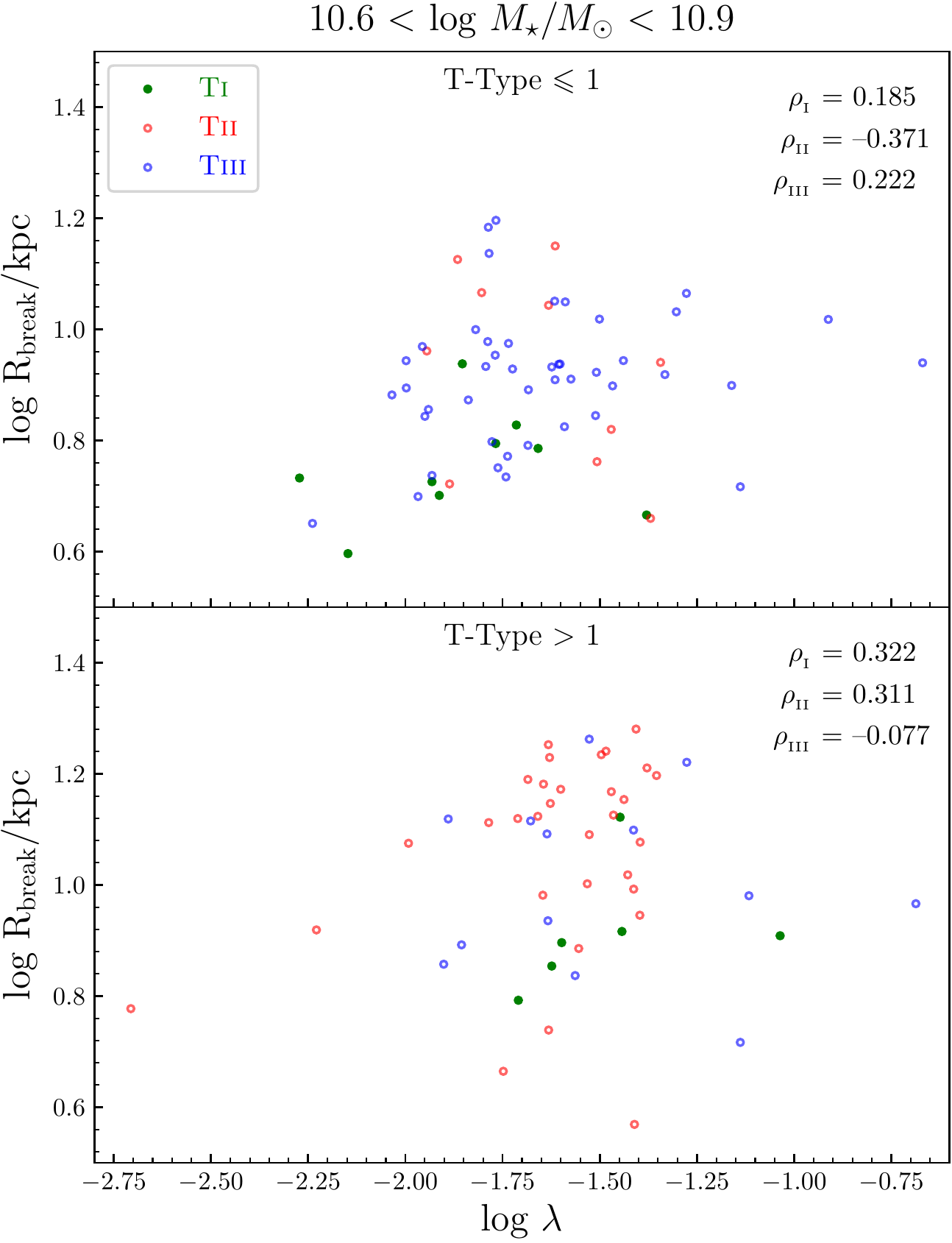}
\caption{
Surface brightness profile break radii of early-type (T-Type $\leqslant$ 1, top panel) and late-type (T-Type $>$ 1, bottom panel) galaxies with stellar masses comparable to the Milky Way are plotted as a function of the spin parameter.\ The $r$-band equivalent scalelengths are used for the spin parameter estimation.\ The Spearman's rank correlation coefficient $\rho$ for different surface brightness profile types is given at the top right corner of each panel.\
}
\label{fig:spin_rbreak}
\end{figure}

\begin{figure}
\centering
\includegraphics[width=0.45\textwidth]{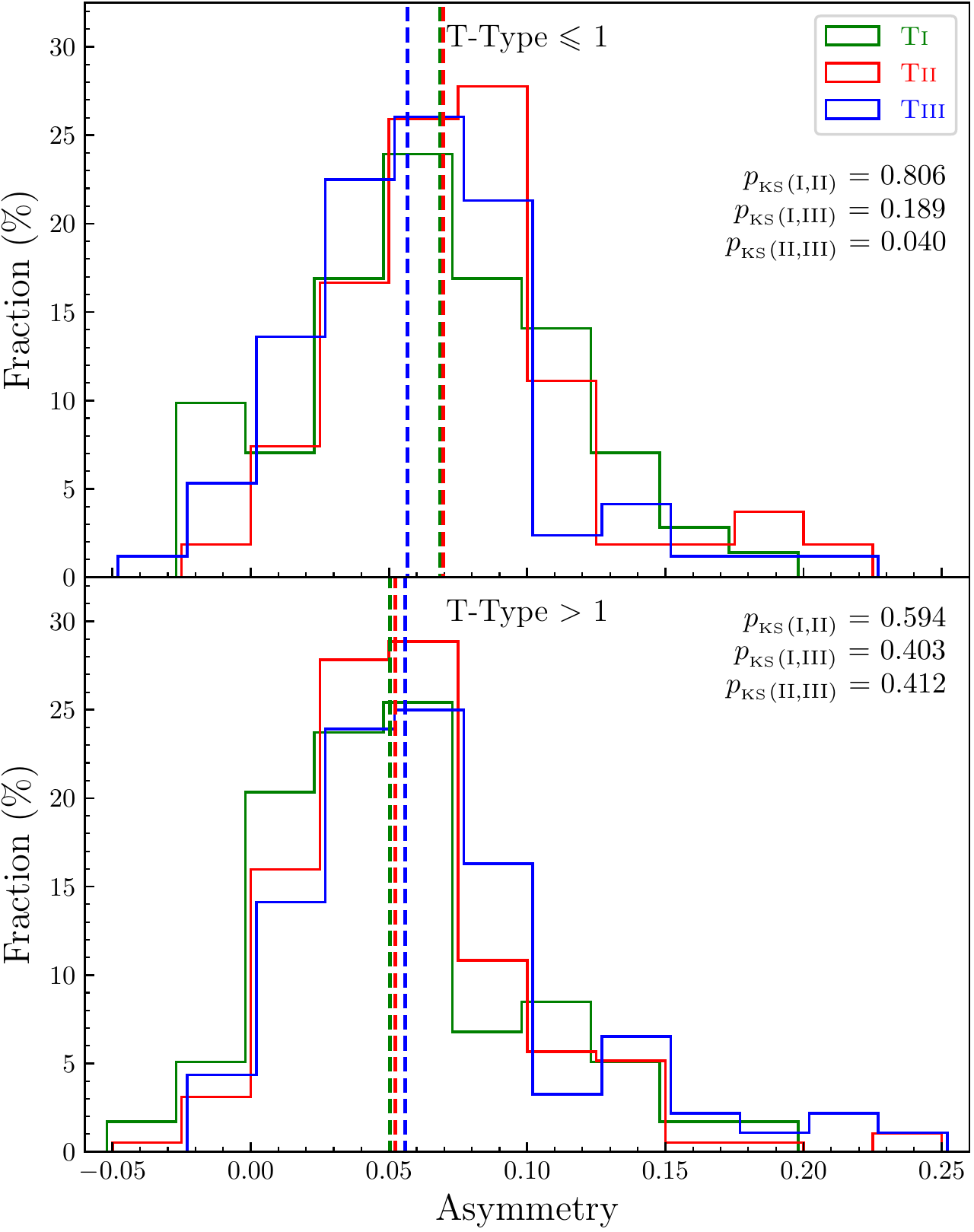}
\caption{
Distribution of the photometric asymmetry parameter for galaxies with different surface brightness profile types.\ The top panel is for early-type galaxies with T-Type $\leqslant$ 1, while the bottom panel is for late-type galaxies with T-Type $>$ 1.\ The vertical dashed lines mark the median asymmetry indices of different subsamples.\ The $p$-values returned by the Kolmogorov-Smirnov test for different surface brightness profile types are given at the top right corner of each panel.}
\label{fig:asymmetry}
\end{figure}

According to the classical disk formation scenario (\citealt{Fall&Efstathiou1980,Dalcanton1997,Mo1998}), the angular momentum of galaxy disks arises in the same way 
as their dark matter halos (i.e., through tidal torques with neighboring structures), and thus the sizes of galaxy disks are scaled with the pre-collapse angular momentum of the host 
halos.\ To add to the plausible relevance of halo angular momentum to disk formation, recent simulations by \cite{Herpich2015} find a correlation between the host halo's initial angular 
momentum and stellar disk profile types, in the sense that {\sc Tii} galaxies formed in halos with the highest angular momentum whereas {\sc Tiii} galaxies formed in halos with the lowest 
angular momentum.\

Here we attempt to make a direct observational test for the relevance of galaxy angular momentum to disk break formation.\ Figure \ref{fig:spinhist} presents the distribution of the 
spin parameter $\lambda$ for disk classifications based on surface brightness profiles.\ The spin parameter $\lambda$ has been commonly used to quantify the angular momentum 
of galaxies in the literature.\ Estimation of $\lambda$ is described in Section \ref{sec:method_spinpars}.\ We divide the sample into early-type (T-Type $\leqslant$ 1) and late-type (T-Type $>$ 1) 
galaxies and show their $\lambda$ distributions separately in the top and bottom panels of the left column in Figures \ref{fig:spinhist}.\ The $p$-values returned by the Kolmogorov-Smirnov 
test for subsamples with different surface brightness profile types are indicated in the figure.\ For early morphological types, there is no significant ($p>0.05$) difference between the 
$\lambda$ distributions of {\sc Ti}, {\sc Tii} and {\sc Tiii} galaxies, but for late-type morphological types, {\sc Ti} galaxies have a significantly broader $\lambda$ distribution and higher 
median $\lambda$ value than do {\sc Tii} and {\sc Tiii} galaxies, with no significant difference between {\sc Tii} and {\sc Tiii} galaxies.\ Given that the halo spin parameter distribution is 
virtually mass independent and has a very mild cosmic evolution \citep[e.g.,][]{Munoz-Cuartas2011}, our results are in conflict with the above-mentioned simulation results of \cite{Herpich2015}.\

Compared to the surface brightness profile, stellar mass profile is probably a more fundamental property to use when exploring the connection with spin parameters.\
We classify the $\Sigma_{\star}$ profiles of our galaxies into three types, in the same way that we do for classifying the surface brightness profiles except that the $\Sigma_{\star}$ 
break radii, if any, are fixed to that of the surface brightness profiles.\ To estimate the spin parameter appropriate for stellar mass distributions, we also calculate the equivalent 
scalelengths of the $\Sigma_{\star}$ profiles, in the same way as for surface brightness profiles.\ The resultant $\lambda$ distributions based on $\Sigma_{\star}$ profiles are 
shown in the right column of Figure \ref{fig:spinhist}.\ There are no significant differences between different $\Sigma_{\star}$ profile types, irrespective of the morphological types.\

In addition to $\lambda$ distributions, it is also of our interest to verify whether there is a correlation between the break radius $\rbreak$ and $\lambda$ for galaxies of given 
masses, as is found in the simulations of \cite{Herpich2015}.\ In Figure \ref{fig:spin_rbreak}, we explore the relation between $\lambda$ and $\rbreak$ for galaxies with stellar 
masses comparable to that of the Milky Way.\ The Spearman's rank correlation coefficient for galaxies with different T-Types and profile types is indicated in Figure \ref{fig:spin_rbreak}.\ 
There is no significant correlation between $\lambda$ and $\rbreak$ for any subsamples shown in Figure \ref{fig:spin_rbreak}, again suggesting that the present-day disk profile 
break is not directly linked to the galaxy spin parameter.\ Our conclusion here does not depend on the choice of stellar mass ranges.

\subsection{Photometric asymmetry of galaxies with different disk profile types}
Tidal disturbances \citep[e.g.,][]{Erwin2005, Watkins2019} and galaxy mergers \citep[e.g.,][]{Younger2007,Borlaff2014} have been invoked as possible formation mechanisms for stellar profile breaks, especially in {\sc Tiii} galaxies.\ Galaxies involved in tidal interactions are expected to have distorted or peculiar stellar structures.\ Photometric asymmetry measurements have been commonly used for automatic recognition of interacting signatures.\ Here we compare the distribution of the photometric shape asymmetry measurements from the SDSS PawlikMorph catalog\footnote{\href{https://data.sdss.org/datamodel/files/MANGA\_MORPHOLOGY/sedmorph/PawlikMorph.html}{https://data.sdss.org/datamodel/files/MANGA\_MORPHOLOGY/\\sedmorph/PawlikMorph.html}} for our galaxies with different surface brightness profile types in Figure \ref{fig:asymmetry}.\ The shape asymmetry parameters are measured using the 8-connected structure detection algorithm to enhance the detectability of low surface brightness tidal features \citep{Pawlik2016}.

As shown in Figure \ref{fig:asymmetry}, there is no significant difference between the asymmetry parameter distributions of late-type galaxies with different surface brightness types.\ For early-type galaxies, the only (marginally) significant difference is that between {\sc Tii} and {\sc Tiii} galaxies, with a Kolmogorov-Smirnov test $p$-value of 0.04.\ The early-type {\sc Ti} and {\sc Tii} galaxies have slightly larger median asymmetry values than {\sc Tiii} galaxies, which appears to be contrary to the popular thought that mergers or satellite accretion may play a more important role in producing {\sc Tiii} profiles \citep[e.g.,][]{Watkins2019}.\

\section{SUMMARY}
\label{sec:summary}

In an effort to probe the origin of surface brightness profile breaks observed in nearby disk galaxies, we have carried out a comparative study of the radial profiles of surface brightness and stellar populations (out to 2.5$\re$) of galaxies with single exponential ({\sc Ti}), down-bending double exponential ({\sc Tii}) and up-bending double exponential ({\sc Tiii}) surface brightness profiles, based on IFU spectroscopic data of 635 disk galaxies selected from the SDSS DR15 MaNGA data release.\ In particular, we use the $\hdeltaf$ index and EW(H$\alpha$) to trace the luminosity-weighted stellar ages, the [MgFe]$^{\prime}$ and Mg/Fe indices to trace stellar  metallicities, and the H$\alpha$ luminosities to trace the recent SFR.\ We also derive stellar mass surface density profiles based on mass-to-light ratios estimated from a full spectrum fitting and galaxy spin parameters $\lambda$ based on rotation curves extracted from the spectroscopic data cubes.\ Our main results and their implications are summarized below.\

\begin{itemize}

\item {\sc Tii} profiles are mainly found in late-type galaxies (T-Type $\gtrsim$ 1), while {\sc Tiii} profiles are mainly found in early-type galaxies (T-Type $\lesssim$ 1).\ This general trend is consistent with previous studies.\ Once dividing our sample into early-type and late-type galaxies, {\sc Ti}, {\sc Tii} and {\sc Tiii} galaxies follow about the same stellar mass-$\rtf$ relation (Figure \ref{fig:massradius}), where $\rtf$ is the $g$-band 25 mag arcsec$^{-2}$ isophotal radius.\ At given stellar masses and T-Types, $\rtf$ corresponds to nearly the same stellar mass surface densities for the three profile types.\  Moreover, the typical radii of surface brightness profile breaks, once normalized by $\rtf$, are about the same for {\sc Tii} and {\sc Tiii} galaxies, with a median $\rbreak$/$\rtf$ ratio $\simeq$ 0.6 (Figure \ref{fig:hisrbreak}).\ The similar mass-size relation for different profile types makes the hypothesis that stellar radial migration alone can transform disk profiles from {\sc Tii} to {\sc Ti} and then to {\sc Tiii} types highly unlikely, because radial migration is expected to gradually increase stellar densities beyond the break radii.\ This conclusion is corroborated by our finding of no significant difference between the stellar metallicity gradients of {\sc Ti} and {\sc Tii} galaxies (Figure \ref{fig:gradinout}).\

\item Galaxies exhibit a diversity of radial profile shapes in the age/metallicity-sensitive spectral features, irrespective of their surface brightness profile types.\ Nevertheless, as one of the most age-sensitive Lick indices, $\hdeltaf$ has a single linear radial profile for a dominant fraction of early-type disk galaxies (T-Type $\leqslant$ 1) while a down-bending radial profile for a dominant fraction of late-type disk galaxies (T-Type $>$ 1), irrespective of the surface brightness profile types (Figure \ref{fig:hisrbreak}).\ The familiar U-shape stellar age profiles, as represented by $\Lambda$-shape $\hdeltaf$ profiles in this paper, are the dominant ones only for late-type {\sc Tii} galaxies. 

\item As a tracer of {\it in situ} star formation intensities, $\Sigma_{\rm H\alpha}$ has a down-bending radial declining for most {\sc Tii} galaxies while an up-bending radial declining for most {\sc Tiii} galaxies, irrespective of T-Types (Figure \ref{fig:gradinout}).\ This provides an unambiguous evidence that abrupt changes of star formation intensities from the inner to outer disks contribute to the formation of both {\sc Tii} and {\sc Tiii} breaks, and at the same time rules out the possibility that a superposition of thin and thick disks with different scalelengths is an important mechanism for forming {\sc Tiii} breaks in our sample.\ Nevertheless, a comparison between the observed distribution of our galaxies and previous simulations \citep{Sanchez-Blazquez2009} on the radial break strengths of $\Sigma_{\rm SFR}$ versus $\Sigma_{\star}$ plane suggests that stellar radial migration plays a significant role in weakening the down-bending $\Sigma_{\star}$ profile breaks produced by {\it in-situ} star formation (Figure \ref{fig:dslope_sfr_mass}).\ According to these simulations, more than half of the stars beyond the break radius have migrated from inner disks.\

\item There is a general correspondence between the break strengths (i.e., outer-minus-inner disk gradient slope differences) of spectral feature and surface brightness profiles for {\sc Tii} galaxies, in the sense that stronger down-bending surface brightness profile breaks correspond to stronger down-bending $\hdeltaf$ (and EW(H$\alpha$)) profile breaks and stronger up-bending [MgFe]$^{\prime}$ (and Mg/Fe) profile breaks.\ No such correlations are found for {\sc Tiii} galaxies (Figure \ref{fig:bkstrength_comp1}).\ {\sc Tiii} galaxies have close to zero median break strengths in $\hdeltaf$ and EW(H$\alpha$) profiles and much weaker median up-bending breaks in [MgFe]$^{\prime}$ and Mg/Fe profiles than {\sc Tii} galaxies.\ The lack of correlation for {\sc Tiii} galaxies may imply that stellar radial migration does not play a major role in shaping the up-bending disk profiles, as is evidenced by the good correspondence between the break strengths of $\Sigma_{\star}$ and optical stellar light profiles.\

\item Contrary to predictions from some recent simulations which invoke galaxy spin parameter to explain the formation of different disk profile types \citep[e.g.,][]{Herpich2015}, we do not find significant differences between $\lambda$ distributions of {\sc Ti}, {\sc Tii} and {\sc Tiii} galaxies, nor do we find significant correlations between $\lambda$ and break radius at given stellar masses (Figures \ref{fig:spinhist}, \ref{fig:spin_rbreak}).\

\item There are no significant differences between photometric asymmetries of different profile types, suggesting that environmental disturbances or satellite accretion in the recent past do not play an important role in the formation of surface brightness profile breaks (Figure \ref{fig:asymmetry}).

\end{itemize}

Above all, we conclude that {\sc Tii} surface brightness breaks are primarily formed by abrupt drops in star formation intensities beyond the break radii, but stellar radial migration plays a significant role in weakening the resultant down-bending $\Sigma_{\star}$ profile breaks.\ It is obvious that {\it in situ} star formation also contributes to the formation of {\sc Tiii} breaks.\ Our finding that {\sc Tiii} galaxies (especially at later 
morphological types) have steeper inner-disk star formation radial gradient slopes than {\sc Ti} and {\sc Tii} galaxies implies that an enhancement of star formation intensities at smaller 
radii, instead of at larger radii, is the primary mechanism for shaping {\sc Tiii} profiles in out sample.\

\begin{acknowledgements}

We are grateful to the anonymous referee for very helpful comments that improve the paper.\ This work is supported by the National Key R\&D Program of China (2017YFA0402600, 2017YFA0402702), the B-type Strategic Priority Program of the Chinese Academy of Sciences (XDB41000000), the NSFC grant (Nos. 11973038, 11973039 and 11421303), and the CAS Pioneer Hundred Talents Program.

Funding for the Sloan Digital Sky Survey IV has been provided by the Alfred P. Sloan Foundation, the U.S. Department of Energy Office of Science, and the Participating Institutions. SDSS-IV acknowledges support and resources from the Center for High-Performance Computing at the University of Utah. The SDSS web site is www.sdss.org. SDSS-IV is managed by the Astrophysical Research Consortium for the Participating Institutions of the SDSS Collaboration including the Brazilian Participation Group, the Carnegie Institution for Science, Carnegie Mellon University, the Chilean Participation Group, the French Participation Group, Harvard-Smithsonian Center for Astrophysics, Instituto de Astrof\'isica de Canarias, The Johns Hopkins University, Kavli Institute for the Physics and Mathematics of the Universe (IPMU) / University of Tokyo, the Korean Participation Group, Lawrence Berkeley National Laboratory, Leibniz Institut f\"ur Astrophysik Potsdam (AIP), Max-Planck-Institut f\"ur Astronomie (MPIA Heidelberg), Max-Planck-Institut f\"ur Astrophysik (MPA Garching), Max-Planck-Institut f\"ur Extraterrestrische Physik (MPE), National Astronomical Observatories of China, New Mexico State University, New York University, University of Notre Dame, Observat\'ario Nacional / MCTI, The Ohio State University, Pennsylvania State University, Shanghai Astronomical Observatory, United Kingdom Participation Group, Universidad Nacional Aut\'onoma de M\'exico, University of Arizona, University of Colorado Boulder, University of Oxford, University of Portsmouth, University of Utah, University of Virginia, University of Washington, University of Wisconsin, Vanderbilt University, and Yale University.

\end{acknowledgements}

\vspace{2pt}
\bibliography{reference}
\bibliographystyle{aasjournal}

\restartappendixnumbering

\appendix

\section{Representative examples of radial profiles of surface brightness, spectral features and $\Sigma_{\star}$ for {\sc Ti} and {\sc Tiii} galaxies}

\begin{figure*}[b]
\centering
\includegraphics[width=0.99\textwidth]{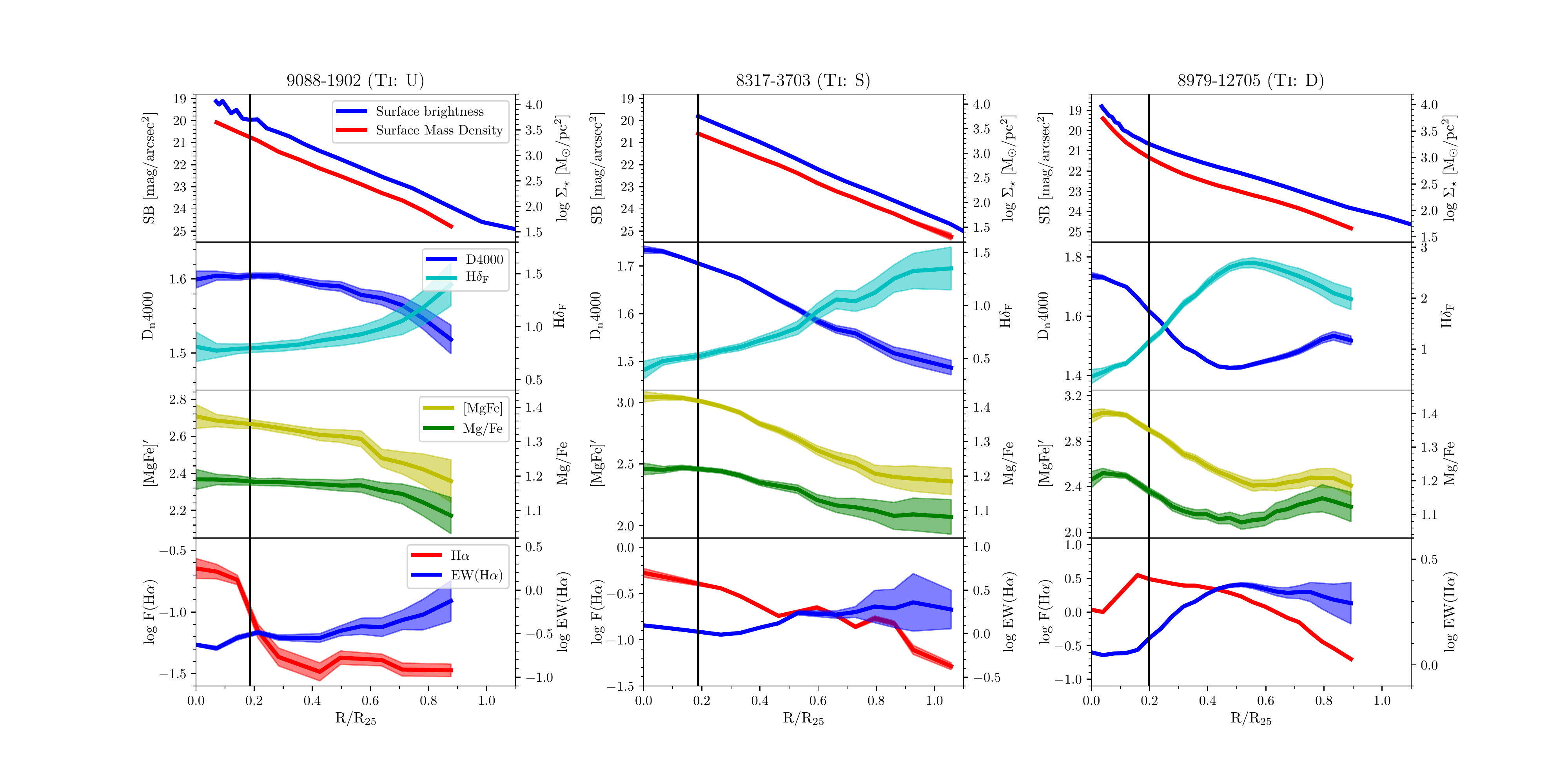}
\caption{Same as Figure \ref{fig:examplet2}, but for examples of {\sc Ti} surface brightness profiles.}
\label{fig:examplet1}
\end{figure*}

\begin{figure*}
\centering
\includegraphics[width=0.99\textwidth]{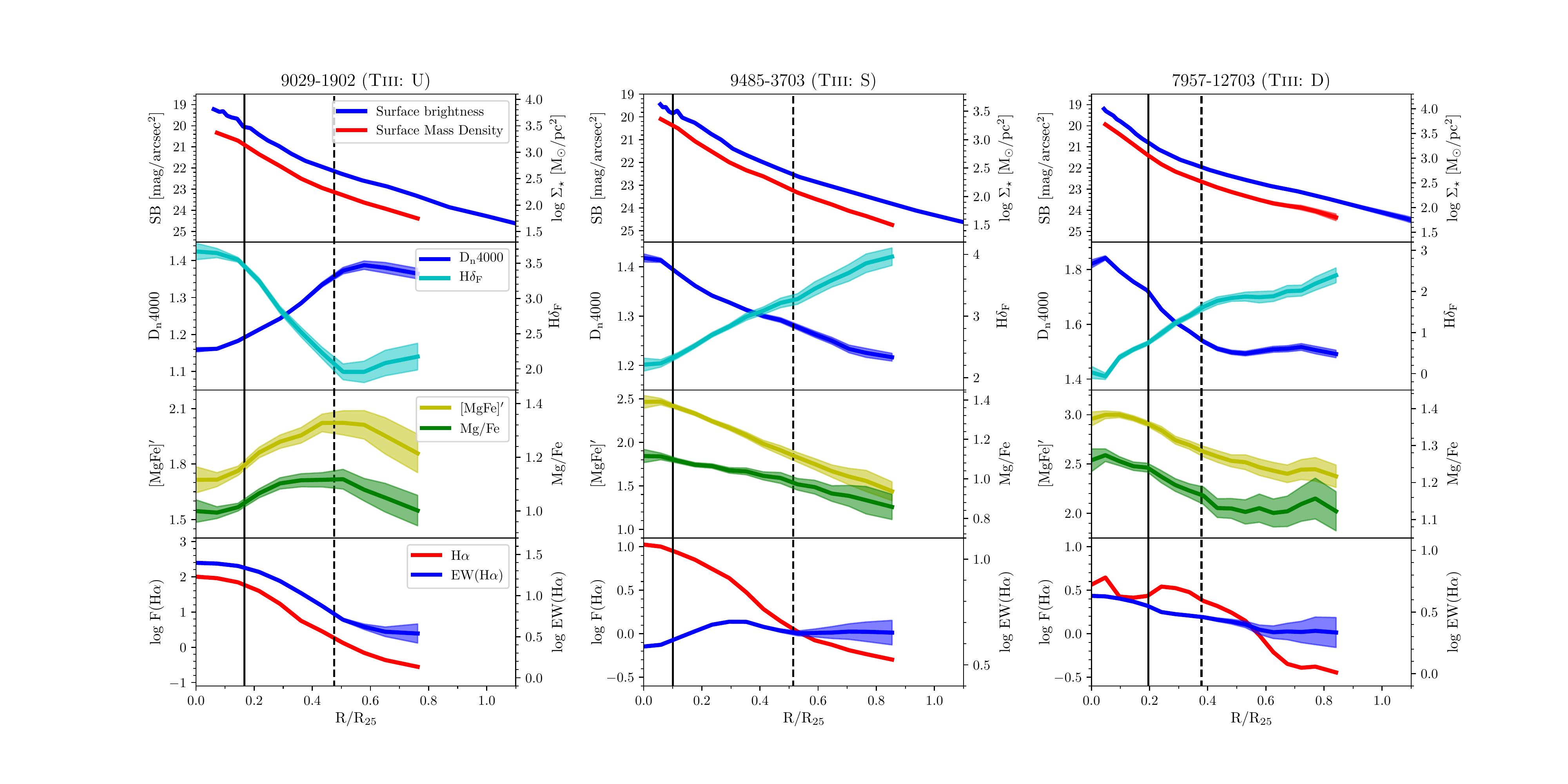}
\caption{Same as Figure \ref{fig:examplet2}, but for examples of {\sc Tiii} surface brightness profiles.}
\label{fig:examplet3}
\end{figure*}

\end{document}